\documentclass[10pt]{iopart}
\usepackage{graphicx}
\eqnobysec

\newcommand{\beq}{\begin{equation}}
\newcommand{\beqa}{\begin{eqnarray}}
\newcommand{\eeq}{\end{equation}}
\newcommand{\eeqa}{\end{eqnarray}}
\newcommand{\abs}[1]{\left\vert#1\right\vert}
\newcommand{\bin}[2]{{\displaystyle{{\displaystyle#1}\choose{\displaystyle#2}}}}
\newcommand{\comport}[2]{\mathrel{\mathop{#1}\limits_{#2}^{}}}
\renewcommand{\d}{{\rm d}}
\newcommand{\dis}[1]{{\overline{#1}}}
\renewcommand{\e}{{\rm e}}
\newcommand{\eps}{\varepsilon}
\newcommand{\erfc}{\mathop{\rm erfc}\nolimits}
\newcommand{\frad}[2]{\displaystyle{\displaystyle#1\over\displaystyle#2}}
\newcommand{\gam}{\gamma}
\newcommand{\hk}{{\hat k}}
\newcommand{\ii}{{\rm i}}
\renewcommand{\max}{{\rm max}}
\renewcommand{\min}{{\rm min}}
\newcommand{\p}{\psi}
\newcommand{\ph}{\phi}
\newcommand{\var}[1]{\mathop{{\rm var}}\nolimits\left(#1\right)}
\newcommand{\w}{\omega_0}
\newcommand{\z}{\zeta}
\newcommand{\prob}[1]{\mathop{\rm Prob}\nolimits\{#1\}}

\newcommand{\F}{{\cal F}}
\renewcommand{\H}{{\cal H}}
\newcommand{\I}{{\cal I}}
\newcommand{\M}{{\cal M}}
\newcommand{\Q}{{\cal Q}}
\newcommand{\R}{{\cal R}}
\newcommand{\T}{{\cal T}}
\newcommand{\X}{{\cal X}}

\begin{document}

\title[Unusual electronic properties of zigzag graphene nanoribbons]
{Unusual electronic properties of clean and disordered zigzag graphene nanoribbons}

\author{J M Luck$^1$ and Y Avishai$^2$}

\address{$^1$ Institut de Physique Th\'eorique, URA 2306 of CNRS, CEA Saclay,
91191~Gif-sur-Yvette cedex, France}
\address{$^2$ Department of Physics and the Ilse Katz Center for Nano-Science,
Ben-Gurion University, Beer-Sheva 84105, Israel}
\eads{\mailto{jean-marc.luck@cea.fr},\mailto{yshai@bgu.ac.il}}

\begin{abstract}
We revisit the problem of electron transport
in clean and disordered zigzag graphene nanoribbons,
and expose numerous hitherto unknown peculiar properties of these systems at zero energy,
where both sublattices decouple because of chiral symmetry.
For clean ribbons,
we give a quantitative description of the unusual power-law dispersion
of the central energy bands and of its main consequences,
including the strong divergence of the density of states near zero energy,
and the vanishing of the transverse localization length of the corresponding edge states.
In the presence of off-diagonal disorder,
which respects the lattice chiral symmetry,
all zero-energy localization properties are found to be anomalous.
Recasting the problem in terms of coupled Brownian motions
enables us to derive numerous asymptotic results by analytical means.
In particular the typical conductance $g_N$ of a disordered sample of width $N$
and length $L$ is shown to decay as $\exp(-C_Nw\sqrt{L})$,
for arbitrary values of the disorder strength $w$,
while the relative variance of $\ln g_N$ approaches a non-trivial constant $K_N$.
The dependence of the constants $C_N$ and $K_N$ on the ribbon width $N$ is predicted.
From the mere viewpoint of the transfer-matrix formalism,
zigzag ribbons provide a case study with many unusual features.
The transfer matrix describing propagation through one unit cell of a clean ribbon
is not diagonalizable at zero energy.
In the disordered case, we encounter non-trivial random matrix products
such that all Lyapunov exponents vanish identically.
\end{abstract}


\maketitle

\section{Introduction}
\label{intro}

Ever since the experimental discovery of graphene, the list of its amazing
properties increases without let-off~\cite{GN,roche,gui,Sarma}.
These range from technological aspects,
graphene being a light and strong material that conducts charge and heat and can be
integrated into many devices, to fundamental features,
such as Dirac cones near zero energy,
relativistic physics at $v_F\ll c$, Klein paradox, zero-energy Landau level,
and many others.

Investigating electronic properties in the bulk,
i.e., for an infinite 2D sheet of graphene,
is relatively simple because translation invariance holds in both directions.
But an infinite system is an idealization,
and, sooner or later, one faces the task of analyzing electronic properties of
finite (or semi-infinite) systems, and especially systems with boundaries.
This is important if one is interested, for example, in the integration of
graphene into an electrical circuit or any other device.
The difference of electronic properties between a bulk infinite graphene sheet
and a system with boundaries such as a nanoribbon is dramatic.
In particular the presence of boundaries may completely modify
the spectrum and the structure of its Dirac points.
The energy dispersion is indeed known to be intimately related with the geometry of
the boundaries~\cite{roche,gui,Sarma}.

It therefore comes as no surprise that, since the early days of graphene,
an extensive research has been devoted to electronic properties of systems
with boundaries in general, and to nanoribbons in
particular~\cite{zz,fwn,nf,bf,me,pg,kh,ab,eskh,wty,wsn,dum,GM,bdj,dbj,kag}.
Nanoribbons are obtained by cutting
a stripe out of the 2D graphene sheet along two parallel lines,
such that translation invariance is maintained along the stripe.
The main possible edge structures are zigzag, armchair and bearded.
Electronic properties of graphene ribbons with zigzag edges
have been shown to exhibit several peculiar features at zero energy.
At this special energy, right at the band center,
both sublattices decouple because of chiral symmetry,
and the system exhibits flat bands and exponentially localized
edge states~\cite{roche,fwn,nf,me,kh,wty,wsn,GM,kag}.
Zero energy is also usually the Fermi energy, if the system is neither doped or gated.

In the present work we revisit this problem in a self-contained way,
considering successively clean and disordered zigzag ribbons
within the one-particle tight-binding framework.
This study will unveil many novel features of these systems at zero energy.
The unusual power-law dispersion of the central bands
provides the leading thread which unites all these properties.

The setup of this paper is as follows.
Section~\ref{clean} is devoted to clean zigzag ribbons.
In Section~\ref{clean:band}
we give a fully self-contained account of their band structure and edge states.
We emphasize the power-law dispersion of the two central bands
around zero energy, and several of its consequences.
For a ribbon of width $N$, i.e., consisting of $N$ coupled chains,
this dispersion law reads
$E\approx\pm Q^N$, where $Q=\pi-q$ is the momentum difference
relative to the boundary of the Brillouin zone.
Surprisingly, this unusual dispersion law,
with its exponent equal to the ribbon width $N$,
was previously noticed only in a single recent unpublished work~\cite{GM}.
Section~\ref{clean:states} contains an investigation of generic zero-energy
eigenstates of a semi-infinite ribbon.
The structure of a special zero-energy state
that has a unit amplitude on the upper left corner of the ribbon
reveals a peculiar pattern where the amplitudes build up a Pascal triangle.
In Section~\ref{disorder} we turn
to the investigation of zero-energy properties of zigzag ribbons
in the presence of hopping (off-diagonal) disorder.
Our main motivation for choosing this type of disorder is theoretical:
off-diagonal disorder respects the so-called chiral symmetry between both sublattices.
It may therefore be expected to keep some
of the very peculiar zero-energy features of clean ribbons.
Moreover, hopping disorder might be induced by applying a weak inhomogeneous strain,
as hopping integrals have a strong exponential dependence
on interatomic distances (see~\cite{strain}).
We shall investigate in particular
the distribution of generic zero-energy eigenstates in
Section~\ref{disorder:states}
and of the zero-energy conductance in Section~\ref{disorder:conductance},
using the transfer-matrix approach to the electronic conductance
of quasi-one-dimensional systems, which has been developed in the early days
of mesoscopic physics~\cite{pic1,mps,pic3,mpk,rmp}.
Recasting the problem in terms of coupled Brownian motions
evolving in a fictitious continuous time $\tau=w^2l$ enables us
to derive by analytical means asymptotically exact predictions
for many quantities of interest, for an arbitrary disorder strength $w$.
As a general rule,
all observables related to localization properties are anomalous.
A typical zero-energy wavefunction on a semi-infinite ribbon
exhibits an anomalous sub-exponential growth with the distance $l$
from the end of the ribbon, scaling as $\exp(A_nw\sqrt{l})$ on chain number~$n$.
The typical conductance $g_N$ of a finite disordered ribbon of width~$N$ and length~$L$
exhibits an anomalous sub-exponential decay with $L$,
scaling as $\exp(-C_Nw\sqrt{L})$.
Band-center anomalies in quasi-one-dimensional bipartite systems
with off-diagonal disorder in the tight-binding formalism have been studied long ago,
first in the case of chains~\cite{TC,FL,SJ,IT,JV},
and then for rectangular strips made of $N$ coupled channels
for odd $N$~\cite{strip1,strip2,strip3,strip4}.
In the present situation, the source of anomalous behavior is different.
Unusual behaviour is predicted at zero energy,
irrespective of the number of channels, either even or odd.
Section~\ref{discussion} contains a brief discussion of our main findings.
In~\ref{appa} we employ elementary methods to study zero-energy properties
of a single chain,
while~\ref{appb} is devoted to the statistics of zero-energy eigenstates
at chain number $n=2$.

\section{Clean ribbons}
\label{clean}

\begin{figure}[!ht]
\begin{center}
\includegraphics[angle=-90,width=.45\linewidth]{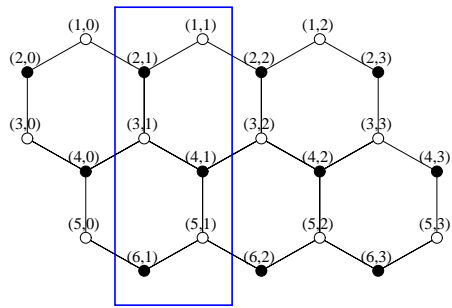}
\caption{\small
A zigzag nanoribbon of width $N=3$.
Blue box: unit cell containing 6 sites, 3 even ones (full symbols)
and 3 odd ones (empty symbols).}
\label{labels}
\end{center}
\end{figure}

A zigzag graphene nanoribbon consists of $N$ coupled chains,
as shown in Figure~\ref{labels} for $N=3$.
It is a periodic array whose unit cell (blue box) contains $2N$ sites.
Sites are conveniently labelled $i=(m,l)$,
where $m=1,\dots,2N$ is the site number within a cell,
while $l$ is the cell number along the ribbon.
The ribbon is bipartite,
with even and odd $m$ corresponding to the two equivalent sublattices,
each cell containing~$N$ even sites and $N$ odd sites.

\subsection{Band structure}
\label{clean:band}

Zigzag ribbons have been known for long to have unusual
electronic properties~\cite{roche,fwn,nf,me,kh,wty,wsn,GM,kag}.
The existence of localized edge states at zero energy
and the flatness of the central bands near the boundaries of the Brillouin zone
are the most salient of these characteristic properties,
which are absent in other geometries, such as armchair ribbons or nanotubes.

Let us begin with a self-contained
direct derivation of the band structure of zigzag ribbons,
without using the two-dimensional band structure of the graphene sheet.
We consider the one-particle tight-binding Hamiltonian
\beq
\H=\sum_{<i,j>}(a^\dag_ia_j+a^\dag_ja_i),
\label{ham}
\eeq
where the sum runs over pairs of nearest neighbors,
and the hopping amplitude has been set to unity.
For an eigenstate at energy $E$, the amplitudes $\p_{m,l}$ obey
\beq
\matrix{
E\p_{4k+1,l}=\p_{4k,l}+\p_{4k+2,l}+\p_{4k+2,l+1},\hfill\cr\ms
E\p_{4k+2,l}=\p_{4k+1,l-1}+\p_{4k+1,l}+\p_{4k+3,l},\hfill\cr\ms
E\p_{4k+3,l}=\p_{4k+2,l}+\p_{4k+4,l-1}+\p_{4k+4,l},\hfill\cr\ms
E\p_{4k+4,l}=\p_{4k+3,l}+\p_{4k+3,l+1}+\p_{4k+5,l}.\hfill
}
\label{tbe}
\eeq

\begin{figure}[!ht]
\begin{center}
\includegraphics[angle=-90,width=.45\linewidth]{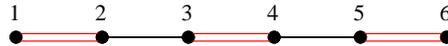}
\caption{\small
Effective dimerized chain corresponding to the ribbon shown in Figure~\ref{labels}.
Single black lines: unit hopping amplitudes.
Double red lines: $\gam$ hopping amplitudes.}
\label{dimer}
\end{center}
\end{figure}

Let us consider an infinitely long ribbon and look for Bloch states of the form
\beq
\matrix{
\p_{4k+1,l}=\ph_{4k+1}\,\e^{\ii ql},\hfill\cr\ms
\p_{4k+2,l}=\ph_{4k+2}\,\e^{\ii q(l-1/2)},\hfill\cr\ms
\p_{4k+3,l}=\ph_{4k+3}\,\e^{\ii q(l-1/2)},\hfill\cr\ms
\p_{4k+4,l}=\ph_{4k+4}\,\e^{\ii ql},\hfill
}
\eeq
where the longitudinal momentum $q$ is in the first Brillouin zone ($\abs{q}\le\pi$).
The transverse amplitudes $\ph_m$ obey the relations
\beq
\matrix{
E\ph_{2n+1}=\ph_{2n}+\gam\ph_{2n+2},\hfill\cr\ms
E\ph_{2n+2}=\gam\ph_{2n+1}+\ph_{2n+3},\hfill
}
\label{ephi}
\eeq
with
\beq
\gam=2\cos\frac{q}{2}.
\label{gamdef}
\eeq
We thus obtain an effective $q$-dependent tight-binding model
on a finite dimerized chain of $2N$ sites~\cite{roche,GM}, as shown in Figure~\ref{dimer}.
The hopping amplitudes originating from vertical bonds (single black lines) equal unity,
whereas those originating from pairs of oblique bonds (double red lines) equal $\gam$.

Looking for a solution to~(\ref{ephi}) of the form
\beq
\matrix{
\phi_{2n+1}=A\,\e^{\ii np},\hfill\cr\ms
\phi_{2n+2}=B\,\e^{\ii np},\hfill
}
\eeq
where $p$ is the transverse momentum,
we obtain the transverse dispersion relation
\beq
E^2=1+2\gam\cos p+\gam^2
\label{dtrans}
\eeq
and the quantization condition
\beq
\sin Np+\gam\sin(N+1)p=0.
\label{qtrans}
\eeq
The above equations~\cite{roche,wsn,GM} possess two kinds of solutions.

\begin{itemize}

\item
{\it Bulk states}, corresponding to real values of $p$, are transversely extended.
Their energies read
\beq
E=\pm\frac{\sin p}{\sin(N+1)p}.
\label{eext}
\eeq

\item
{\it Edge states}, corresponding to complex values of $p$, of the form $p=\pi+\ii\kappa$,
are exponentially localized in the transverse direction
($\ph_{2n+1}\sim\ph_{2n+2}\sim(-1)^n\e^{\pm n\kappa}$).
The existence of localized edge states has been known
for long~\cite{roche,fwn,nf,kh,wty,wsn,GM}.
In the present framework, we have
\beq
\sinh N\kappa=\gam\sinh(N+1)\kappa
\label{kloc}
\eeq
and the corresponding energies read
\beq
E=\pm\frac{\sinh\kappa}{\sinh(N+1)\kappa}.
\label{eloc}
\eeq

\end{itemize}

\begin{figure}[!ht]
\begin{center}
\includegraphics[angle=-90,width=.45\linewidth]{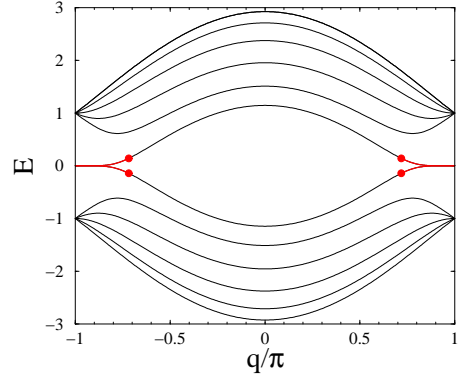}
\caption{\small
Band structure of the zigzag ribbon for $N=6$.
Black: bulk states.
Red: edge states.
Red symbols: transition points ($q=\pm q_N$).}
\label{bands}
\end{center}
\end{figure}

The resulting band structure is shown in Figure~\ref{bands} for $N=6$.
Edge states only contribute to the wings of the central bands.
More precisely, for $\abs{q}<q_N$, with
\beq
\cos\frac{q_N}{2}=\frac{N}{2(N+1)},
\eeq
all the bands correspond to bulk states.
In the wings of the Brillouin zone ($\abs{q}>q_N$),
the two central bands correspond to edge states.
Both expressions~(\ref{eext}) and~(\ref{eloc})
agree to give $E=\pm1/(N+1)$ at the transition points $q=\pm q_N$ (symbols).

The transverse localization length $\xi$ of the edge states
is obtained by taking the $N\to\infty$ limit in~(\ref{kloc}).
We thus obtain~\cite{fwn,nf,kh,wty,wsn,GM}
\beq
\frac{1}{\xi}=\kappa=\ln\frac{1}{\gam}=\ln\frac{1}{2\cos(q/2)}.
\eeq
The localization length $\xi$ diverges at the threshold for the occurrence of edge states,
$q_\infty=\lim q_N=2\pi/3$.
This simple result can be predicted by means of a topological argument
based on the Zak phase~\cite{dum}.

It is worth scrutinizing the vicinity of the boundaries of the Brillouin zone.
Right at the boundaries ($q=\pm\pi$), the hopping amplitude $\gam$ vanishes,
and so the dimerized chain splits up into $(N-1)$ dimers
and two isolated atoms at the endpoints.
Now, setting $\abs{q}=\pi-Q$, with $Q$ small, we have
\beq
\gam\approx Q,\quad\kappa\approx\ln\frac{1}{Q}.
\label{gkq}
\eeq
The pairs of dimer levels ($E=\pm1$) hybridize to give $2(N-1)$ bands of bulk states,
with $p\approx a\pi/N$, and so
\beq
E\approx\pm\left(1+Q\,\cos\frac{a\pi}{N}\right)\quad(a=1,\dots,N-1).
\eeq
The two atomic levels ($E=0$)
give rise to the central bands of edge states.
Equations~(\ref{eloc}) and~(\ref{gkq}) yield
\beq
E\approx\pm Q^N.
\label{epower}
\eeq

The wings of the central bands therefore obey an unusual power-law dispersion,
whose exponent is the ribbon width $N$, and whose prefactor is exactly unity.
This result, which has far reaching consequences,
seems to have been noticed so far only in~\cite{GM}.

The contribution of the central bands to the density of states per site
therefore diverges as a power law as $E\to0$, for any $N\ge2$,
according to
\beq
\rho(E)\approx\frac{1}{2\pi N}\frac{\d Q}{\d E}
\approx\frac{\abs{E}^{-(N-1)/N}}{2\pi N^2}.
\label{rhopower}
\eeq
This divergence explains the observed unusual behavior~\cite{nf}
of the flatness index~$\eta$, defined as the portion of the central bands
contained in a small energy interval $\abs{E}<\Delta$ around zero energy.
We indeed predict
\beq
\eta\approx N\int_{-\Delta}^\Delta\rho(E)\,\d E\approx\frac{\Delta^{1/N}}{\pi}.
\eeq
Another consequence of the power-law dispersion~(\ref{epower})
is the vanishing of the transverse localization length $\xi$
of the edge states as $E\to0$, according to
\beq
\frac{1}{\xi}\approx\ln\frac{1}{Q}\approx\frac{1}{N}\ln\frac{1}{\abs{E}}.
\label{xipower}
\eeq

\subsection{Generic zero-energy eigenstates}
\label{clean:states}

The above analysis of the band structure demonstrates that zero energy
is very peculiar,
as it coincides with the endpoints of the two central bands of edge states,
with the power-law dispersion~(\ref{epower}).
On an infinitely long ribbon,
the Hamiltonian~(\ref{ham}) has only two zero-energy Bloch states,
one living on each sublattice:
\beq
\p_{1,l}=(-1)^l\quad\hbox{and}\quad\p_{2N,l}=(-1)^l.
\label{bloch}
\eeq
These states are strictly confined at the edges,
in agreement with the fact that $\xi\to0$ as $E\to0$ (see~(\ref{xipower})).

It is therefore natural to wonder what generic zero-energy eigenstates look like,
besides the above Bloch states which live on an infinitely long,
translationally invariant ribbon.
The goal of this section is to investigate this question in detail.

As expected from chiral symmetry,
the tight-binding equations~(\ref{tbe}) on both sublattices decouple at zero energy.
Furthermore, they can be recast into the recursive form
\beq
\matrix{
\p_{4k+2,l+1}=-\p_{4k,l}-\p_{4k+2,l},\hfill\cr\ms
\p_{4k+4,l+1}=-\p_{4k+2,l+1}-\p_{4k+4,l},\hfill
}
\label{teven}
\eeq
\beq
\matrix{
\p_{4k+3,l+1}=-\p_{4k+3,l}-\p_{4k+5,l},\hfill\cr\ms
\p_{4k+1,l+1}=-\p_{4k+3,l+1}-\p_{4k+1,l}.\hfill
}
\label{todd}
\eeq

For definiteness, we consider a semi-infinite ribbon starting with cell number $l=0$
and extending infinitely far to the right.
Furthermore, we focus our attention onto the even sublattice.
The recursion relations~(\ref{teven}) can be solved as follows.

\begin{itemize}

\item
For $n=2$, we have $\p_{2,l+1}=-\p_{2,l}$, and so
\beq
\p_{2,l}=(-1)^l\p_{2,0}.
\eeq
The initial condition $\p_{2,0}$ is arbitrary.

\item
For $n=4$, we have $\p_{4,l+1}=-\p_{2,l+1}-\p_{4,l}$, and so
\beq
\p_{4,l}=(-1)^{l+1}(l+1)\p_{2,0}.
\eeq
In particular $\p_{4,0}=-\p_{2,0}$.

\item
For $n=6$, we have $\p_{6,l+1}=-\p_{4,l}-\p_{6,l}$, and so
\beq
\p_{6,l}=(-1)^l\Bigl(\p_{6,0}-\frac{1}{2}\,l(l+1)\p_{2,0}\Bigr).
\eeq
The initial condition $\p_{6,0}$ is arbitrary.

\item
For $n=8$, we have $\p_{8,l+1}=-\p_{6,l+1}-\p_{8,l}$, and so
\beq
\p_{8,l}=(-1)^{l+1}\Bigl((l+1)\p_{6,0}-\frac{1}{6}\,l(l+1)(l+2)\p_{2,0}\Bigr).
\eeq
In particular $\p_{8,0}=-\p_{6,0}$.

\end{itemize}

The above results show the structure
of a generic zero-energy eigenstate on the even sublattice of a semi-infinite ribbon.
The arbitrary initial values
are those at the leftmost column of sites, i.e., $\p_{4k+2,0}$.
The amplitudes grow as various powers of the distance $l$ along the ribbon.
The fastest growth is proportional to the initial condition~$\p_{2,0}$.

It is worth considering in more detail the special eigenstate
obtained if the only non-zero initial condition at $l=0$ is $\p_{2,0}=1$.
This special zero-energy eigenstate is remarkable.
Its amplitudes build up a Pascal triangle.
In other words, they can be expressed in terms of binomial coefficients:
\beq
\matrix{
\p_{4k+2,l}=(-1)^{k+l}\bin{k+l}{2k},\hfill\cr\ms
\p_{4k+4,l}=(-1)^{k+l+1}\bin{k+l+1}{2k+1}.\hfill
}
\label{bins}
\eeq

\begin{figure}[!ht]
\begin{center}
\includegraphics[angle=-90,width=.45\linewidth]{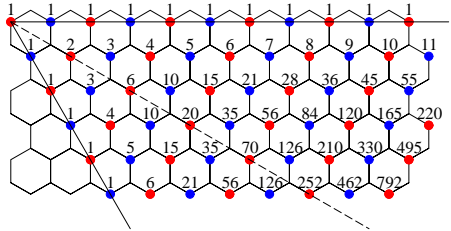}
\caption{\small
Special zero-energy eigenstate on the even sublattice
of a semi-infinite ribbon for $N=6$.
Red symbols: positive amplitudes.
Blue symbols: negative amplitudes.
Full lines: boundaries of the eigenstate.
Dashed line: symmetry axis.}
\label{pascal}
\end{center}
\end{figure}

A similar construction of a symmetric zero-energy eigenstate
induced by a single defect has been given in~\cite{dbj}.
This special eigenstate is shown in Figure~\ref{pascal} for $N=6$.
Positive (resp.~negative) amplitudes are shown as red (resp.~blue) symbols.
Full lines show the boundaries of the eigenstate, where amplitudes equal $\pm1$.
The dashed line shows the symmetry axis of the eigenstate ($l=3k$ or $3k+1$).
The amplitudes of the special eigenstate grow as successive powers of $l$:
\beq
\abs{\p_{2n,l}}\approx\frac{l^{n-1}}{(n-1)!}.
\label{power}
\eeq
The fastest growth is observed for $n=N$, i.e., at the lower edge of the ribbon.
The eigenstate thus becomes more and more strongly localized at the lower edge
as one goes deeper and deeper into the semi-infinite ribbon.
This observation goes hand in hand
with the strict confinement of the zero-energy Bloch states~(\ref{bloch}) at the edges,
and with the fact that $\xi\to0$ as $E\to0$ (see~(\ref{xipower})).

A few words about the relation of the above eigenstates
to the usual notion of a normalizable wavefunction are in order.
Generic generic eigenstates solve the zero-energy one-boundary problem,
where we impose the amplitudes at one end of a semi-infinite ribbon
and propagate the solution by the tight-binding equations
or, equivalently, by the transfer matrix.
The eigenstates thus constructed are not normalizable in general.
This should not be a surprise, as
they do not solve the standard (two-boundaries) Sturm-Liouville problem.
In the presence of off-diagonal disorder,
generic eigenstates will be shown to grow subexponentially with distance $l$
(see~(\ref{abres})).

\section{Disordered ribbons}
\label{disorder}

We now turn to the study of electronic properties of zigzag ribbons
with off-diagonal disorder.
The corresponding Hamiltonian reads
\beq
\H=\sum_{<i,j>}t_{ij}(a^\dag_ia_j+a^\dag_ja_i),
\label{hamd}
\eeq
where the sum runs over pairs of nearest neighbors,
and the hopping amplitudes~$t_{ij}$ are independent random variables.
This type of disorder respects the lattice chiral symmetry.
It may therefore be expected to keep some
of the peculiar zero-energy features of clean ribbons.
Tight-binding Hamiltonians with off-diagonal disorder
on some other bipartite structures~\cite{IT},
including chains~\cite{TC,FL,SJ,JV} and strips~\cite{strip1,strip2,strip3,strip4},
have been shown to exhibit anomalous localization at zero energy.
Throughout this section, we focus our attention onto zero-energy properties.
We parametrize the hopping amplitudes as
\beq
t_{ij}=\e^{\eps_{ij}},
\eeq
where the $\eps_{ij}$ are independent random variables,
drawn from an unspecified symmetric probability distribution
with zero mean and variance $w^2$:
$\dis{\eps_{ij}}=0$, $\var{\eps_{ij}}=w^2$.
The positive parameter~$w$ thus represents the disorder strength.\footnote{Hereafter
the bar denotes an average over the disorder,
and we use the notation $\var{X}=\dis{X^2}-\dis{X}^2$
for the disorder variance of a quantity $X$.}

\begin{figure}[!ht]
\begin{center}
\includegraphics[angle=-90,width=.45\linewidth]{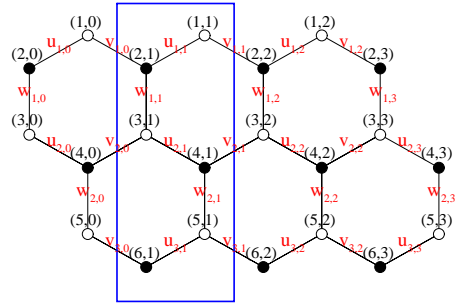}
\caption{\small
Notations for the hopping amplitudes attached to various bonds.}
\label{hoppings}
\end{center}
\end{figure}

For convenience, we introduce the notations
\beq
u_{n,l}=\exp\bigl(\eps_{n,l}^{(u)}\bigr),\quad
v_{n,l}=\exp\bigl(\eps_{n,l}^{(v)}\bigr),\quad
w_{n,l}=\exp\bigl(\eps_{n,l}^{(w)}\bigr)
\eeq
for the hopping amplitudes attached to various types of bonds,
as shown in Figure~\ref{hoppings}.

\subsection{Zero-energy eigenstates}
\label{disorder:states}

In this section we investigate the behavior
of generic zero-energy eigenstates of the disordered Hamiltonian~(\ref{hamd}),
following the line of thought of Section~\ref{clean:states}.
At zero energy, the disordered tight-binding equations
can be recast into a recursive form generalizing~(\ref{teven}) and~(\ref{todd}):
\beq
\matrix{
v_{2k+1,l}\p_{4k+2,l+1}=-w_{2k,l}\p_{4k,l}-u_{2k+1,l}\p_{4k+2,l},\hfill\cr\ms
u_{2k+2,l+1}\p_{4k+4,l+1}=-w_{2k+1,l+1}\p_{4k+2,l+1}-v_{2k+2,l}\p_{4k+4,l},\hfill
}
\label{tevend}
\eeq
\beq
\matrix{
v_{2k+2,l}\p_{4k+3,l+1}=-u_{2k+2,l}\p_{4k+3,l}-w_{2k+2,l}\p_{4k+5,l},\hfill\cr\ms
u_{2k+1,l+1}\p_{4k+1,l+1}=-w_{2k+1,l+1}\p_{4k+3,l+1}-v_{2k+1,l}\p_{4k+1,l}.\hfill
}
\label{toddd}
\eeq

Here again, we consider a semi-infinite ribbon starting with unit number $l=0$,
and extending infinitely far to the right,
and we focus our attention onto the even sublattice.
It is advantageous to reformulate the recursion relations~(\ref{tevend})
as difference equations.
To do so, let us introduce the products
\beq
U_{n,l}=\prod_{m=0}^{l-1} u_{n,m},\quad
V_{n,l}=\prod_{m=0}^{l-1} v_{n,m}.
\eeq
Setting
\beq
\matrix{
\p_{4k+2,l}=(-1)^{k+l}\frad{U_{2k+1,l}}{V_{2k+1,l}}S_{4k+2,l},\hfill\cr\ms
\p_{4k+4,l}=(-1)^{k+l+1}\frad{V_{2k+2,l}}{U_{2k+2,l+1}}S_{4k+4,l},\hfill
}
\label{pd}
\eeq
we obtain
\beq
\matrix{
S_{4k+2,l+1}-S_{4k+2,l}=\frad{V_{2k,l}V_{2k+1,l}}{U_{2k,l+1}U_{2k+1,l+1}}
\,w_{2k,l}S_{4k,l},\hfill\cr\ms
S_{4k+4,l+1}-S_{4k+4,l}=\frad{U_{2k+1,l+1}U_{2k+2,l+1}}{V_{2k+1,l+1}V_{2k+2,l+1}}
\,w_{2k+1,l+1}S_{4k+2,l+1}.\hfill
}
\label{xd}
\eeq
These difference equations can again be solved recursively.

\begin{itemize}

\item
For $n=2$, we have
\beq
\p_{2,l}=(-1)^l\frad{U_{1,l}}{V_{1,l}}\p_{2,0},
\eeq
i.e., $S_{2,l}=\p_{2,0}$,
where the initial condition $\p_{2,0}$ is arbitrary.

\item
For $n=4$, we have
\beq
\p_{4,l}=(-1)^{l+1}\frad{V_{2,l}}{U_{2,l+1}}S_{4,l},
\eeq
with
\beq
S_{4,l}=\sum_{m=0}^{l-1}\frad{U_{1,m}U_{2,m}}{V_{1,m}V_{2,m}}\,w_{1,m}\p_{2,0}.
\eeq
In particular $S_{4,0}=w_{1,0}\p_{2,0}$ and $\p_{4,0}=-(w_{1,0}/u_{2,0})\p_{2,0}$.

\end{itemize}

It is again worth considering the special zero-energy eigenstate
on the even sublattice for which the initial condition is $\p_{m,0}=\delta_{m,2}$.
In this situation, it is clear from their recursive definition
that all the quantities $S_{2n,l}$ are positive.
As a consequence, the amplitudes $\p_{2n,l}$ of the special eigenstate
on a disordered ribbon have exactly the same signs as expressions~(\ref{bins}),
irrespective of the realization of disorder,
i.e., of the draw of the hopping rates $t_{ij}$.

The analysis of the growth law of the magnitudes $\abs{\p_{2n,l}}$ goes as follows.
Beginning with $n=2$, we consider the quantity
\beq
X_{1,l}=\ln\abs{\p_{2,l}}
=\ln\frad{U_{1,l}}{V_{1,l}}
=\sum_{m=0}^{l-1}\left(\eps_{1,m}^{(u)}-\eps_{1,m}^{(v)}\right).
\eeq
The rightmost side is the sum of $2l$ independent random variables
with zero mean and variance $w^2$.
We have therefore
\beq
\dis{\ln\abs{\p_{2,l}}}=0,\quad
\var{\ln\abs{\p_{2,l}}}=2w^2l.
\label{p2}
\eeq
For a large distance $l$, by the central limit theorem,
$X_{1,l}$ is asymptotically a Gaussian variable with zero mean and variance $2w^2l$.
More generally, for arbitrary $k$, all the quantities
\beq
X_{2k+1,l}=\ln\frad{U_{2k+1,l}}{V_{2k+1,l}},\quad
X_{2k+2,l}=\ln\frad{V_{2k+2,l}}{U_{2k+2,l}}
\label{xkdef}
\eeq
are asymptotically independent Gaussian variables with zero mean and variance $2w^2l$.

It is advantageous to recast~(\ref{pd}) and~(\ref{xd}) in terms of the latter variables.
We thus obtain
\beq
\matrix{
\abs{\p_{4k+2,l}}=\e^{X_{2k+1,l}}\,S_{4k+2,l},\hfill\cr\ms
\abs{\p_{4k+4,l}}=\e^{X_{2k+2,l}}\,\frad{1}{u_{2k+2,l}}S_{4k+4,l},\hfill
}
\label{newpd}
\eeq
and
\beq
\matrix{
S_{4k+2,l+1}-S_{4k+2,l}=\e^{-X_{2k+1,l}}\,\frad{w_{2k,l}}{u_{2k+1,l}}\abs{\p_{4k,l}},\hfill\cr\ms
S_{4k+4,l+1}-S_{4k+4,l}=\e^{-X_{2k+2,l+1}}\,w_{2k+1,l+1}\abs{\p_{4k+2,l+1}}.\hfill
}
\label{newxd}
\eeq

For large distances ($l\gg1$),
the Gaussian variables $X_{n,l}$,
which enter~(\ref{newpd}) and~(\ref{newxd}) exponentially, are typically large.
They are also slowly varying with distance $l$,
in the sense that the differences $X_{n,l+1}-X_{n,l}$ have finite variances $2w^2$.
These observations justify the following simplifying steps.

\begin{enumerate}

\item
It is legitimate to use a continuum approximation,
replacing the discrete distance~$l$ by the fictitious continuous time
\beq
\tau=w^2l.
\eeq
The quantities $X_{n,l}$ become independent Gaussian processes $X_n(\tau)$,
known as {\it Brow\-nian motions}, such that
\beq
\dis{X_n(\tau)}=0,\quad
\var{X_n(\tau)}=2\tau.
\label{xtau}
\eeq
The connection between off-diagonal disorder at zero energy and random walks
(or Brownian motions, which are nothing but the continuum limit thereof)
has been known for long in the case of a single chain~\cite{TC,FL,SJ,JV}.

\item
Within this continuum framework,
consistently discarding all prefactors of order unity,
(\ref{newpd}) and~(\ref{newxd}) can be merged into the integral recursions
\beq
\abs{\p_{2n,l}(\tau)}\approx\e^{X_n(\tau)}
{\displaystyle\int_0^\tau\e^{-X_n(\tau')}\abs{\p_{2n-2,l}(\tau')}\d\tau'}.
\label{p2ntau}
\eeq
The latter equations can be solved recursively
in terms of the $X_n(\tau)$, as
\beq
\matrix{
\abs{\p_{2,l}(\tau)}\approx\e^{X_1(\tau)},\hfill\cr\ms
\abs{\p_{4,l}(\tau)}\approx\e^{X_2(\tau)}
{\displaystyle\int_0^\tau\e^{X_1(\tau')-X_2(\tau')}\d\tau'},\hfill\cr\ms
\abs{\p_{6,l}(\tau)}\approx\e^{X_3(\tau)}
{\displaystyle\int_0^\tau\e^{X_2(\tau')-X_3(\tau')}\d\tau'}\hfill\cr\ms
{\hskip 50pt}
\times{\displaystyle\int_0^{\tau'}\e^{X_1(\tau'')-X_2(\tau'')}\d\tau''},\hfill
}
\label{p2tau}
\eeq
and so on.

\item
The growth of the amplitudes $\abs{\p_{2n,l}}$
can be estimated by evaluating the nested integrals
entering~(\ref{p2tau}) by the saddle-point method,
i.e., by looking for the times $\tau',\tau'',\dots$
which maximize the integrands.
We thus obtain
\beq
\ln\abs{\p_{2n,l}}\approx M_n(\tau)+X_n(\tau),
\label{pln}
\eeq
where the $M_n(\tau)$ are non-trivial random processes,
defined by the recursion relation
\beq
M_n(\tau)=\comport{\max}{0\le\tau'\le\tau}
(M_{n-1}(\tau')+X_{n-1}(\tau')-X_n(\tau')),
\label{mrec}
\eeq
with $M_1(\tau)=0$.
The processes $M_n(\tau)$ inherit the diffusive scaling
of the Brownian motions $X_n(\tau)$ which generate them.
We have therefore in particular
\beq
\dis{M_n(\tau)}=A_n\sqrt\tau,\quad
\var{M_n(\tau)+X_n(\tau)}=B_n\tau,
\eeq
where the amplitudes $A_n$ and $B_n$ are constants
which depend only on the chain number $n$.

\end{enumerate}

\begin{figure}[!ht]
\begin{center}
\includegraphics[angle=-90,width=.45\linewidth]{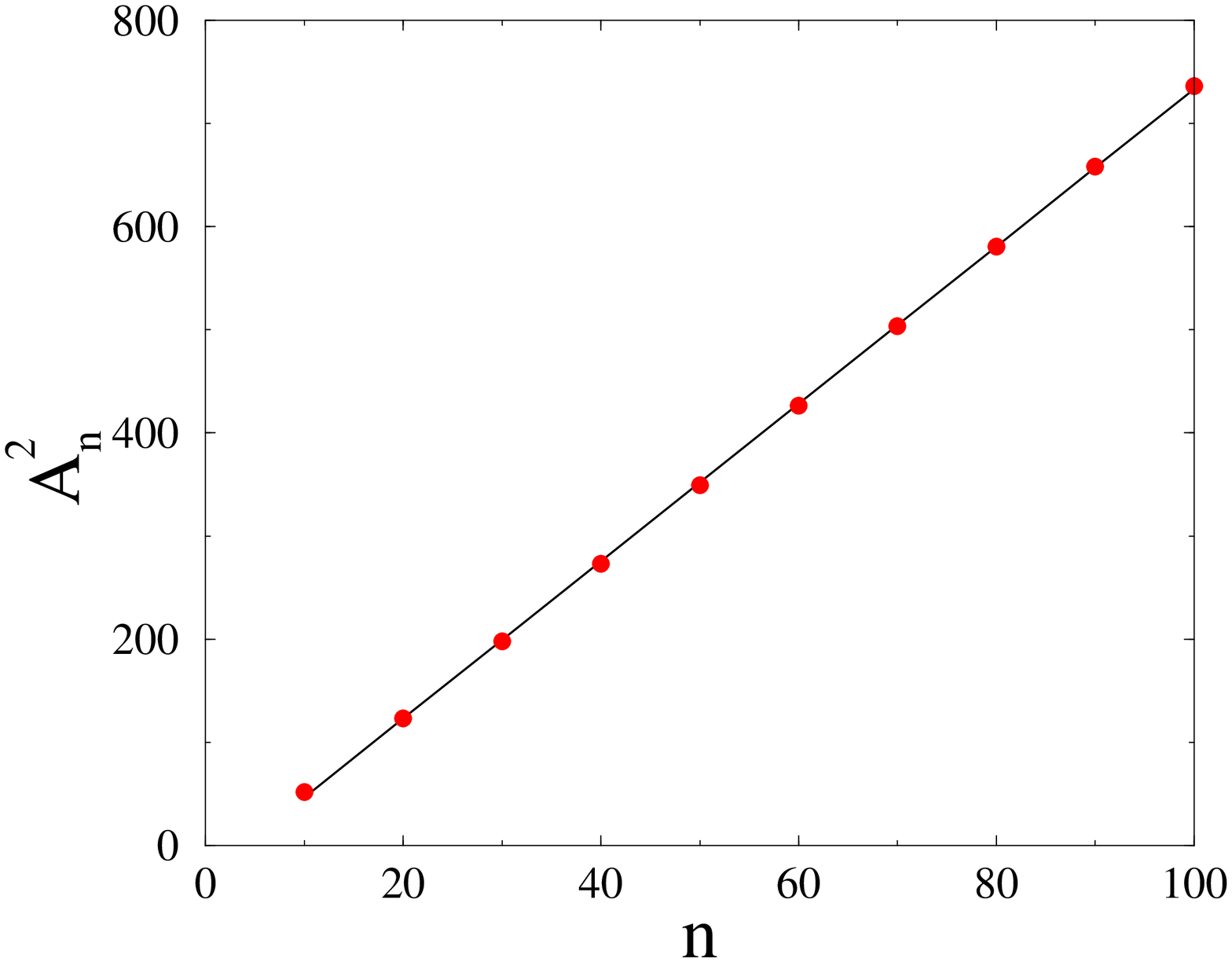}
\includegraphics[angle=-90,width=.45\linewidth]{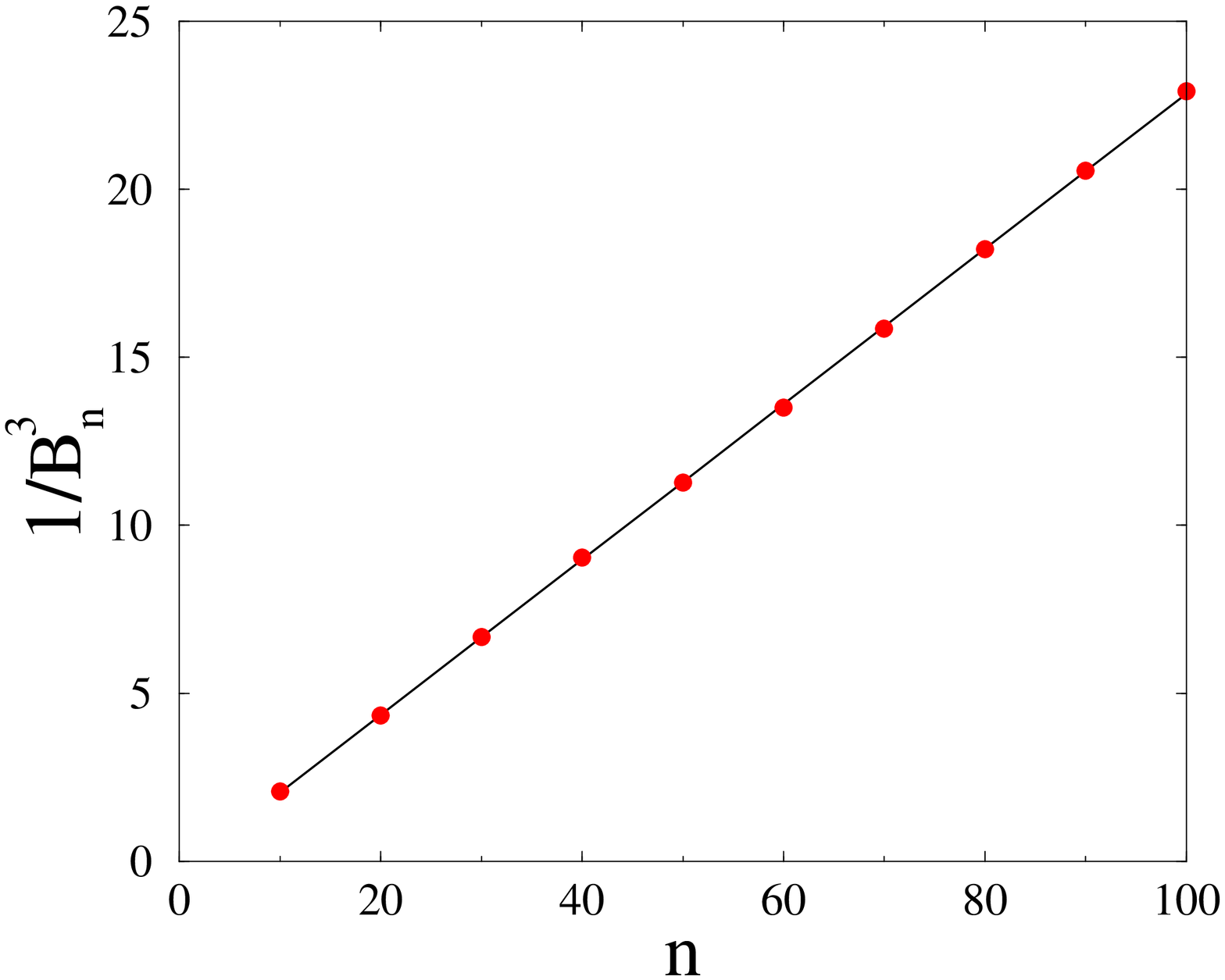}
\caption{\small
Plots of $A_n^2$ and $1/B_n^3$ against chain number $n$.
Straight lines: least-square fits with respective slopes 7.63 and 0.231.}
\label{anbn}
\end{center}
\end{figure}

We are thus left with the asymptotic growth laws
\beq
\dis{\ln\abs{\p_{2n,l}}}\approx A_n w\sqrt{l},\quad
\var{\ln\abs{\p_{2n,l}}}\approx B_n w^2l.
\label{abres}
\eeq
These formulas are the main result of this section.
They hold for all values of the disorder strength $w$.
We thus predict that a generic zero-energy eigenstate
grows subexponentially with distance~$l$ on a disordered ribbon.
This growth follows a power law on a clean ribbon (see~(\ref{power})).
We shall come back in the discussion to the connection
between the growth laws~(\ref{power}) and~(\ref{abres})
and the transfer-matrix formalism.
We have not succeeded in evaluating exactly the constants $A_n$ and $B_n$,
except in the following two cases.
For $n=1$, the result~(\ref{p2}) yields the simple results
\beq
A_1=0,\quad B_1=2.
\eeq
For $n=2$, the problem boils down to the distribution
of the maximum of a single Brownian motion (see~\ref{appb}).
We thus obtain
\beq
A_2=\sqrt{\frac{8}{\pi}},\quad B_2=4-\frac{8}{\pi}.
\label{ab2}
\eeq
For higher $n$, we have obtained numerical values of $A_n$ and $B_n$
by means of a direct simulation of the coupled random processes
$X_n(\tau)$ and $M_n(\tau)$.
Figure~\ref{anbn} shows plots of $A_n^2$ and $1/B_n^3$
against chain number $n$, up to $n=100$.
The good quality of the linear fits strongly suggests
that $A_n$ and $B_n$ scale for large $n$ as
\beq
A_n\approx2.8\sqrt{n},\quad B_n\approx1.6\,n^{-1/3}.
\label{absca}
\eeq

\begin{figure}[!ht]
\begin{center}
\includegraphics[angle=-90,width=.45\linewidth]{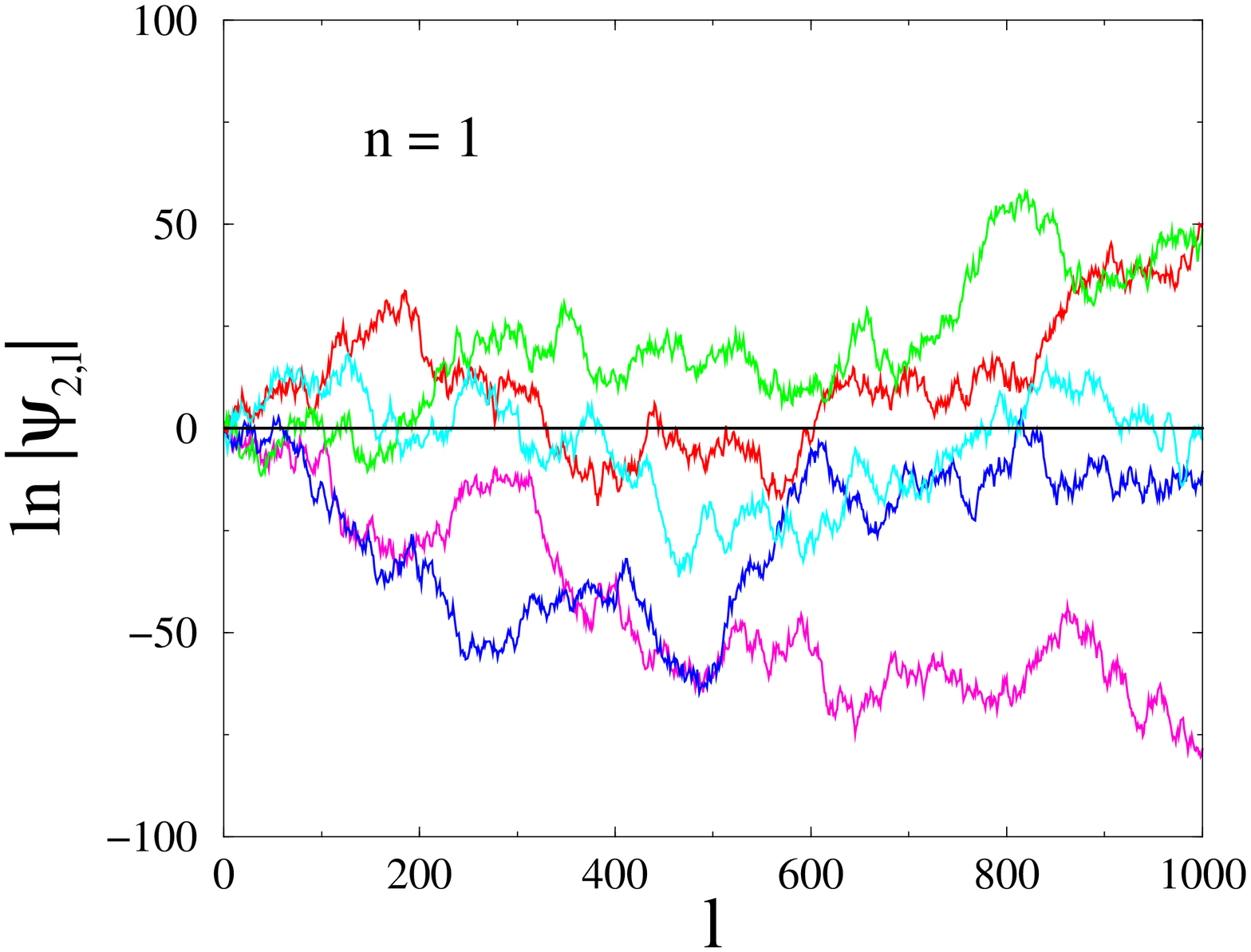}
\includegraphics[angle=-90,width=.45\linewidth]{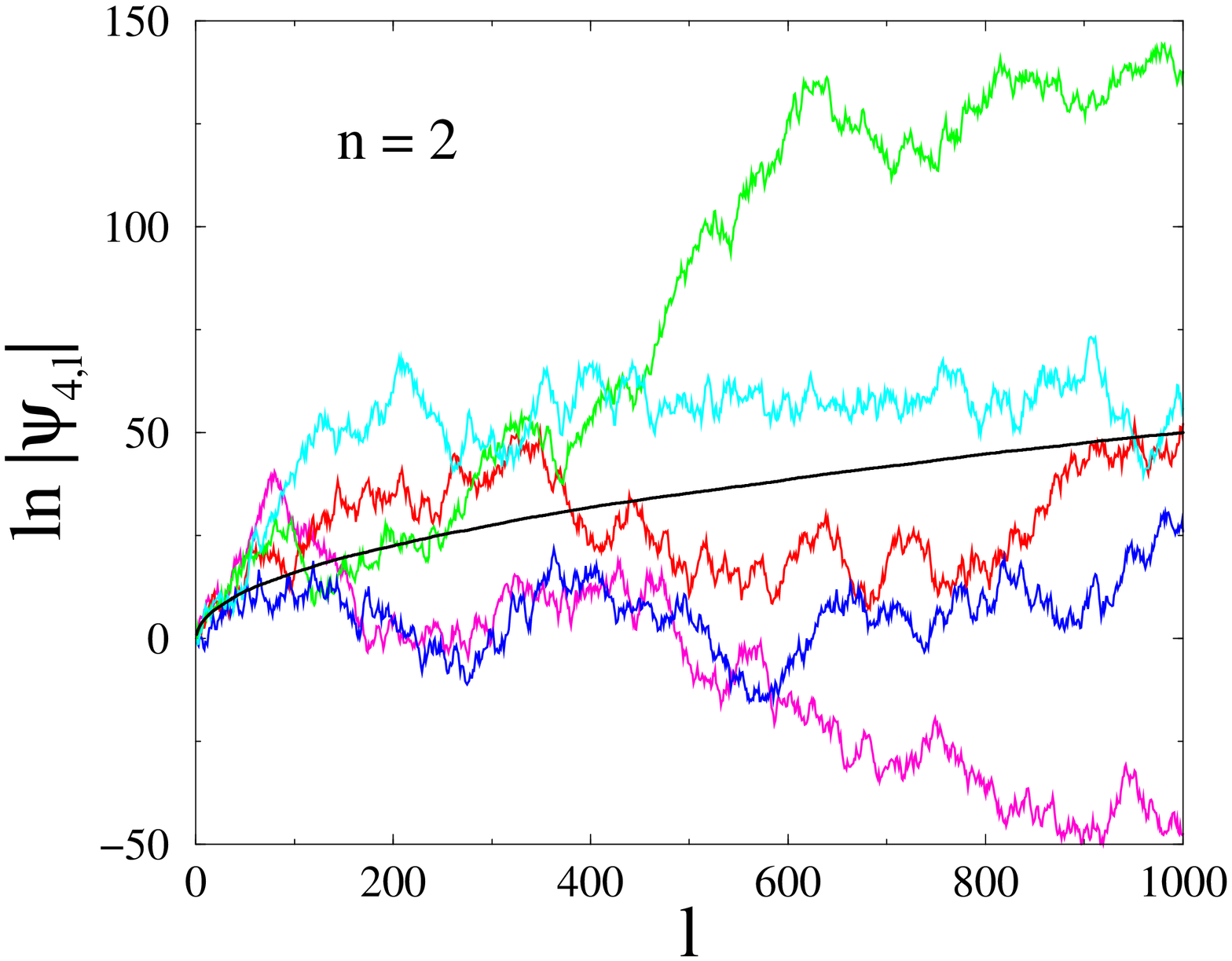}

\includegraphics[angle=-90,width=.45\linewidth]{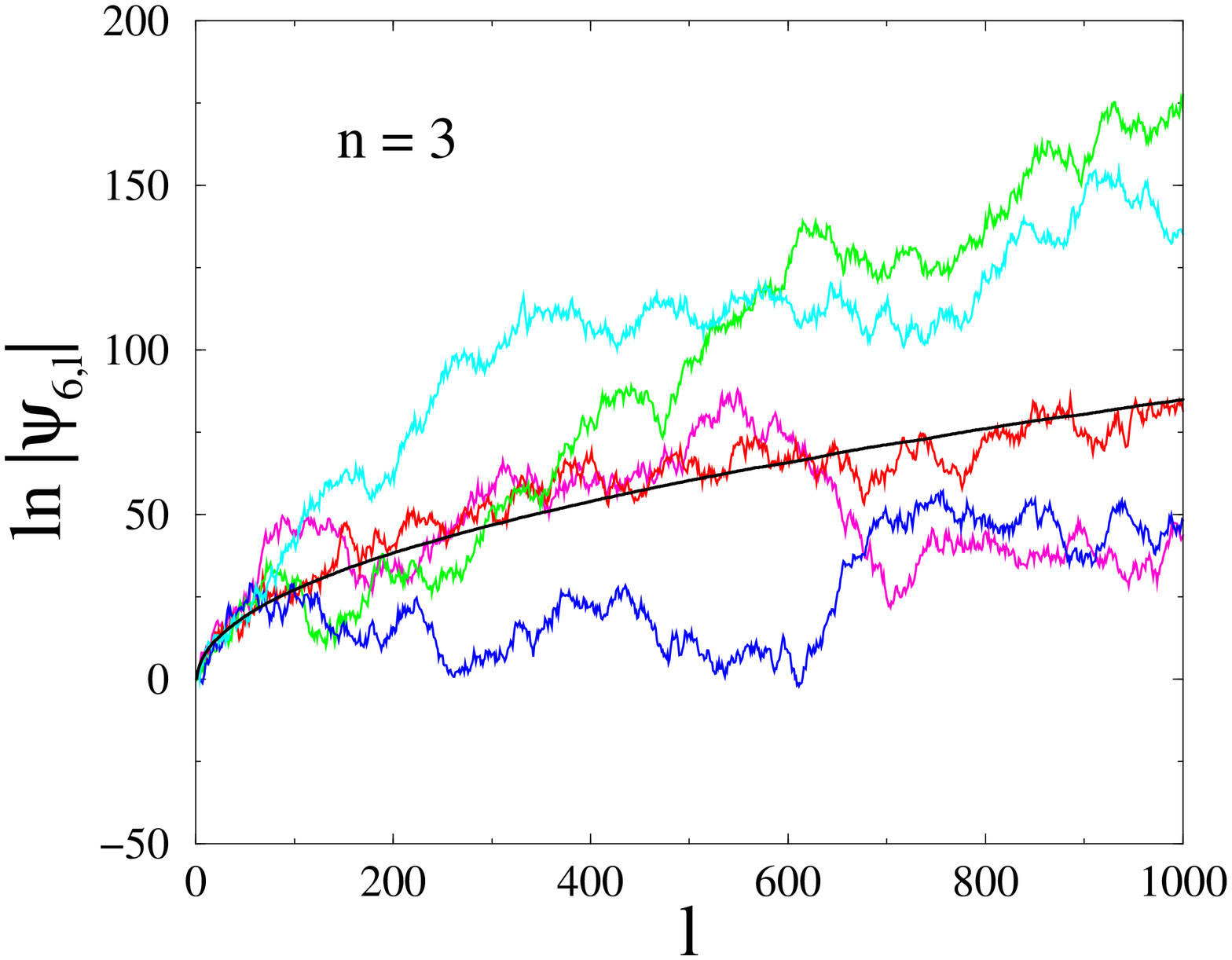}
\includegraphics[angle=-90,width=.45\linewidth]{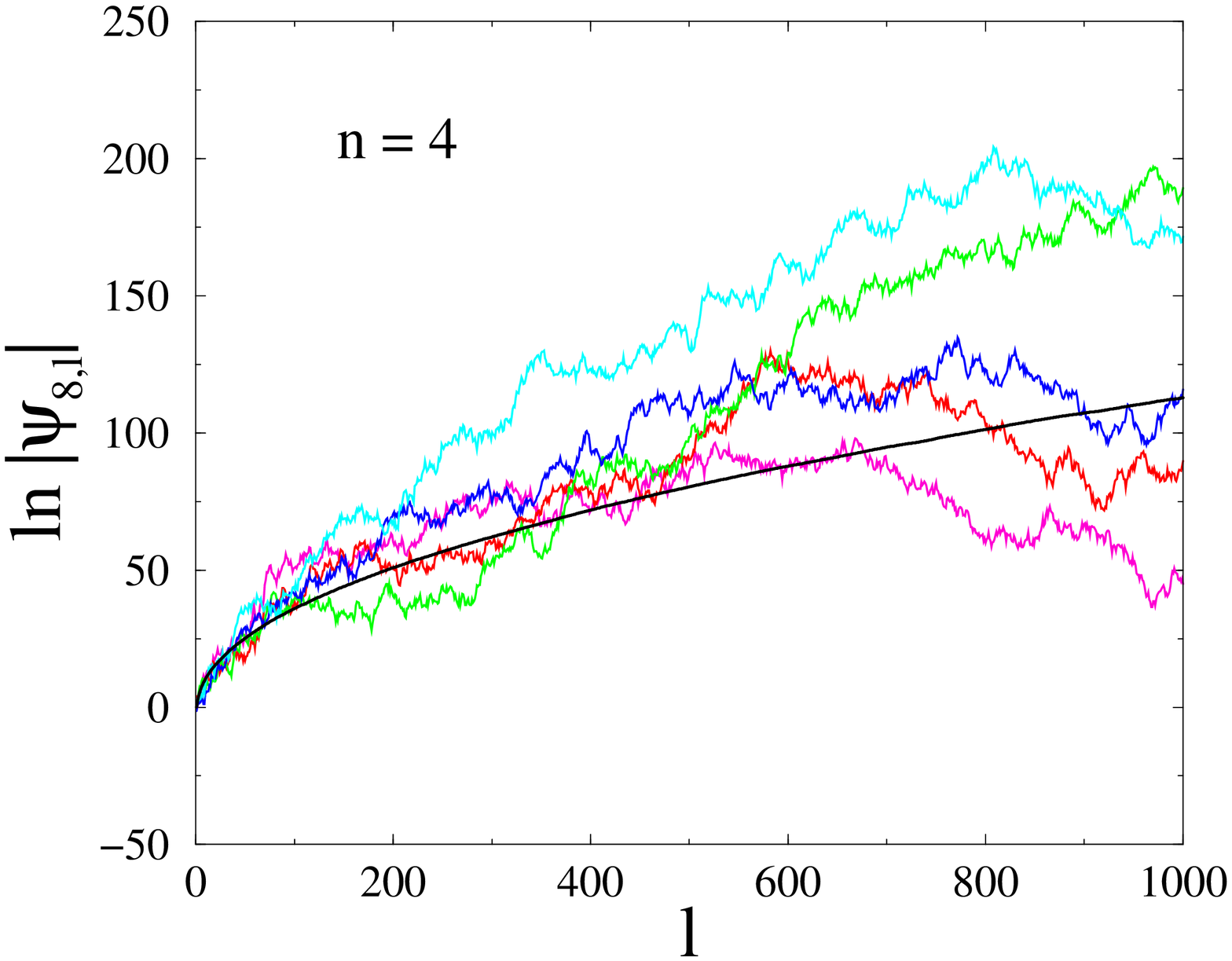}
\caption{\small
Logarithmic plots of $\abs{\p_{2n,l}}$ against distance $l$ along each chain,
for the special zero-energy eigenstate on the even sublattice
of semi-infinite disordered ribbons of width $N=4$, with a disorder strength $w=1$.
Colors code for five different realizations of disorder.
Smooth lines: averages over many samples,
growing asymptotically as $A_n\sqrt{l}$ (see~(\ref{abres})).}
\label{x}
\end{center}
\end{figure}

In order to illustrate the above results,
Figure~\ref{x} shows the special zero-energy eigenstate on the even sublattice
of semi-infinite disordered ribbons of width $N=4$.
The hopping rates are drawn from
the uniform distribution in the interval $-\sqrt3\le\eps\le\sqrt3$,
corresponding to a disorder strength $w=1$.
Colors code for five different realizations of disorder.
The four panels show plots
of $\ln\abs{\p_{2n,l}}$ against the distance~$l$ along each chain.
Smooth lines show averages of the plotted quantities over many samples,
growing asymptotically as $A_n\sqrt{l}$ (see~(\ref{abres})).
For $n=1$, the tracks behave as Brownian motions and have zero average.
For higher $n$ (other panels), averages increase with $n$
(in agreement with the slow growth of $A_n$ with $n$),
while individual tracks show less and less dispersion around the averages
(in agreement with the slow fall-off of $B_n$ with~$n$).

\subsection{Zero-energy conductance}
\label{disorder:conductance}

In this section we investigate the zero-energy conductance
of a disordered zigzag ribbon sample of length $L$ and width $N$,
connected to two leads consisting of semi-infinite clean ribbons of the same width.
We shall use the transfer-matrix approach
to the electronic conductance~\cite{pic1,mps,pic3,mpk,rmp}.
Let us start with a brief reminder of this formalism.
The $2N\times2N$ transfer matrix $T_E$
describing the propagation across {\it one unit cell} of a clean ribbon at energy~$E$
is obtained by recasting the tight-binding equations~(\ref{tbe}) into the form
\beq
\p_{m,l+1}=\sum_{n=1}^{2N}(T_E)_{m,n}\p_{n,l}.
\eeq
Taking for definiteness the example of $N=2$, we thus obtain
\beq
T_E=\pmatrix{E^2-1&-E&1&-E\cr E&-1&0&0\cr 0&0&-1&E\cr -E&1&-E&E^2-1}.
\eeq
Looking for an eigenvalue of $T_E$ of the form $y=\e^{\ii q}$,
we obtain the dispersion law
\beq
2(1+\cos q)=\gam^2=E^2\pm E,
\eeq
with the notation~(\ref{gamdef}).
The above expression agrees with~(\ref{dtrans}) and~(\ref{qtrans}).
Right at zero energy, the transfer matrix
\beq
T_0=\pmatrix{-1&0&1&0\cr 0&-1&0&0\cr 0&0&-1&0\cr 0&1&0&-1}
\eeq
is not diagonalizable.
Indeed, its characteristic polynomial reads $(y+1)^4$,
whereas the eigenvalue $y=-1$, corresponding to the boundaries of the Brillouin zone
($q=\pm\pi$), has only two independent eigenvectors,
\beq
V_-=\pmatrix{1\cr0\cr0\cr0}
\quad\hbox{and}\quad
V_+=\pmatrix{0\cr0\cr0\cr1},
\eeq
in correspondence with the two zero-energy Bloch states~(\ref{bloch}).
The peculiar structure of $T_0$ is stable by multiplication.
The matrix describing the propagation across a clean ribbon of width $N=2$ and length $L$
at zero energy indeed reads
\beq
T_0^L=(-1)^L\pmatrix{1&0&-L&0\cr 0&1&0&0\cr 0&0&1&0\cr 0&-L&0&1}.
\eeq
The power-law growth of the elements of $T_0^L$,
is due to the feature that $T_0$ is not diagonalizable.
It is also closely related to the power-law growth of generic zero-energy eigenstates
on clean ribbons, investigated in Section~\ref{clean:states}.

Now, let $\T$ be the $2N\times2N$ transfer matrix
describing the propagation across the disordered sample of length $L$ at zero energy.
The structure of the tight-binding equations~(\ref{tevend}) and~(\ref{toddd}) implies
that the non-zero elements of $\T$ are
$\T_{2k,2l}$ for $k\ge l$ and $\T_{2k+1,2l+1}$ for $k\le l$.
For $N=2$, we thus obtain
\beq
\T=\pmatrix{
\T_{1,1}&0&\T_{1,3}&0\cr0&\T_{2,2}&0&0\cr
0&0&\T_{3,3}&0\cr0&\T_{4,2}&0&\T_{4,4}}.
\eeq
The transfer matrix $\T$ reduces to $T_0^L$ in the absence of disorder.
The non-zero elements of both matrices are at the same positions,
irrespective of the disorder realization.

According to the transfer-matrix approach~\cite{pic1,mps,pic3,mpk,rmp},
the dimensionless conductance $g_N$ of the sample
(in units of $e^2/h$ and per spin degree of freedom) reads
\beq
g_N=\sum_{a=1}^N\frac{1}{1+x_a},
\label{gdef}
\eeq
where the subscript $N$ denotes the number of channels
(i.e., half the size of the transfer matrix),
while the number $a=1,\dots,N$ labels the channels,
and the $x_a$ are the eigenvalues of the matrix\footnote{$\I$ denotes
the $2N\times2N$ unit matrix.}
\beq
\X=\frac{1}{4}\left(\M^\dag\M+(\M^\dag\M)^{-1}-2\,\I\right),
\label{xdef}
\eeq
where the matrix
\beq
\M=\R^{-1}\T\R
\label{mdef}
\eeq
is obtained by rotating the transfer matrix $\T$ of the sample
to a basis where the transfer matrix $T_0$ of the leads is diagonal.
In other words, the columns of $\R$ are right eigenvectors of~$T_0$.

In the present situation, we are therefore facing an obstacle,
as the transfer matrix~$T_0$ is not diagonalizable.
This peculiarity is related to the fact
that there are only two Bloch states~(\ref{bloch}) at zero energy.
A natural way of regularizing the problem
consists in considering a small non-zero energy $E$.
The eigenvalues $y_a$ of $T_0$ can then be derived from the
power-law dispersion~(\ref{epower}) of the central bands, i.e., $Q^{2N}=E^2$.
We have therefore
\beq
y_a=\e^{\ii q_a}=-\e^{\ii Q_a}\approx-(1+\ii Q_a),
\label{ta}
\eeq
with
\beq
Q_a=Q_0\,\z^a\quad(a=1,\dots,2N),
\eeq
where the momentum scale $Q_0$ is given by
\beq
Q_0=\abs{E}^{1/N},
\label{q0}
\eeq
while the complex number
\beq
\z=\e^{\ii\pi/N}
\label{zeta}
\eeq
is the first $2N$th root of unity.

The momentum scale $Q_0$ provides a cutoff length below which
zero-energy properties are approximately valid.
A finite ribbon of length $L$ can indeed be expected to exhibit
zero-energy features for $Q_0L\ll1$,
and to be effectively far from zero energy in the opposite regime ($Q_0L\gg1$).

In order to proceed,
it is advantageous to change basis according to the following formulas.

\begin{itemize}

\item
For $N=2K$ even, we set
\beq
\matrix{
\R_{a,2k+1}=(-1)^kr_{a,4k+1}=\w^{2k}\z^{2ka},\hfill\cr\ms
\R_{a,2k+2}=(-1)^{k+1}r_{a,4k+3}=\w^{2k+1}\z^{(2k+1)a},\hfill\cr\ms
\R_{a,2\hk}=\eta(-1)^{k+1}r_{a,4K-4k+2}=\w^{2k-1}\z^{(2\hk-1)a},\hfill\cr\ms
\R_{a,2\hk+1}=\eta(-1)^{k+1}r_{a,4K-4k}=\w^{2k}\z^{2\hk a},\hfill
}
\eeq

\item
For $N=2K+1$ odd, we set
\beq
\matrix{
\R_{a,2k+1}=(-1)^kr_{a,4k+1}=\w^{2k}\z^{2ka},\hfill\cr\ms
\R_{a,2k+2}=(-1)^{k+1}r_{a,4k+3}=\w^{2k+1}\z^{(2k+1)a},\hfill\cr\ms
\R_{a,2\hk+1}=\ii\eta(-1)^k r_{a,4K-4k+4}=\w^{2k-1}\z^{2\hk a},\hfill\cr\ms
\R_{a,2\hk+2}=\ii\eta(-1)^{k+1} r_{a,4K-4k+2}=\w^{2k}\z^{(2\hk+1)a},\hfill
}
\eeq
with the notations $\hk=K+k$, $\eta={\rm sign}(E)$, $\w=\ii Q_0$,
and where $r_{a,m}$ denotes the components of the right eigenvector
associated with the eigenvalue $y_a$
in the original basis, where sites are labelled $m=1,\dots,2N$.

\end{itemize}

The above change of basis makes the eigenvectors look much simpler.
The matrix~$\R$ thus defined indeed factorizes as $\R=\Q\F$,
up to inessential phase conventions that we shall not mention explicitly,
where
\beq
\Q={\rm diag}(1,Q_0,\dots,Q_0^{N-1},1,Q_0,\dots,Q_0^{N-1}),
\eeq
while $\F$ is the matrix of the finite Fourier transform, with elements
\beq
\F_{a,b}=\frac{\z^{ab}}{\sqrt{2N}}\quad(a,b=1,\dots,2N).
\eeq
We have $\F^\dag\F=\I$.
So, in the $E\to0$ limit, the matrix
\beq
\M=\R^{-1}\T\R=\F^\dag\Q^{-1}\T\Q\F
\eeq
is unitarily equivalent to the matrix
\beq
\lim_{Q_0\to0}\Q^{-1}\T\Q={\rm diag}(\T_{1,1}\dots\T_{2N,2N}).
\label{matlim}
\eeq
Therefore, in the $E\to0$ limit,
the eigenvalues of $\M$ coincide with the diagonal elements $\T_{m,m}$.
The eigenvalues of $\M^\dag\M$ then read $\abs{\T_{m,m}}^2$.
The first correction to the limit~(\ref{matlim})
is proportional to the momentum scale $Q_0$ (see~(\ref{q0})).

The zero-energy conductance of a disordered ribbon is therefore
entirely dictated by the diagonal elements
$\T_{m,m}$ of the sample transfer matrix $\T$.
The latter matrix elements can be extracted
from the recursion relations~(\ref{tevend}),~(\ref{toddd}).
Using the notation~(\ref{xkdef}), we get
\beqa
&&\abs{\T_{4k+1,4k+1}}=\frac{u_{2k+1,0}}{u_{2k+1,L}}\,\e^{-X_{2k+1,L}},
\nonumber\\
&&\abs{\T_{4k+2,4k+2}}=\e^{X_{2k+1,L}},
\nonumber\\
&&\abs{\T_{4k+3,4k+3}}=\e^{-X_{2k+2,L}},
\nonumber\\
&&\abs{\T_{4k+4,4k+4}}=\frac{u_{2k+2,0}}{u_{2k+2,L}}\,\e^{X_{2k+2,L}}.
\eeqa
In the continuum limit introduced in Section~\ref{disorder:states},
the above expressions simplify to
\beq
\abs{\T_{2n-1,2n-1}}\approx\e^{-X_n(\tau)},\quad
\abs{\T_{2n,2n}}\approx\e^{X_n(\tau)}.
\label{tsca}
\eeq

We recall that the $X_n(\tau)$ are independent Brownian motions such that
$\dis{X_n^2(\tau)}=2\tau$, with $\tau=w^2L$.
Using~(\ref{tsca}), we thus obtain the following universal expression
for the dimensionless zero-energy conductance of a long disordered sample of length $L$:
\beq
g_N=\sum_{n=1}^N\frac{1}{\cosh^2 X_n(\tau)}.
\label{gsum}
\eeq
In the case of a single chain ($N=1$), the formula
\beq
g=\frac{1}{\cosh^2 X(\tau)}
\eeq
is exact for any finite sample
and can be derived by elementary means (see~(\ref{gchain})).

Our prediction~(\ref{gsum}) can therefore be stated as follows:
the zero-energy conductance $g_N$ of a long ribbon
is the sum of the conductances of~$N$ independent chains.
This result seems at first sight hardly compatible with the existence
of only two zero-energy Bloch states~(\ref{bloch}) on an infinitely long clean ribbon.
The power-law dispersion law~(\ref{epower})
however shows that all channels are marginally open at zero energy,
resolving thus the apparent contradiction.

The distribution of the conductance $g_N$
therefore reads\footnote{We use the notation $f(g)$
for the distribution (probability density) of $g$,
as a shorthand for $f_g(g)$,
whenever there is no ambiguity.}
\beq
f(g_N)=\underbrace{f(g)*\cdots*f(g)}_{\hbox{$N$ times}},
\eeq
where stars denote convolution products,
while the distribution $f(g)$ of the con\-duc\-tance of a single chain
is derived in~(\ref{fgchain}).
The above result is however not very useful for practical purposes.
More explicit predictions can be made for short and long samples.

For a weak enough disorder,
such that $w^2L\ll1$, the result~(\ref{gsum}) simplifies as
\beq
g_N\approx N-\sum_{n=1}^NX_n^2(\tau).
\label{quadra}
\eeq
The distribution of the conductance is therefore peaked
around the maximal ballistic value $(g_N=N$), as should be:
\beq
f(g_N)\approx\frac{(N-g_N)^{(N-2)/2}\,\e^{-(N-g_N)/(4w^2L)}}{\Gamma(N/2)(4w^2L)^{N/2}}.
\label{gbal}
\eeq
We have in particular
\beq
\dis{g_N}\approx N(1-2w^2L),\quad\var{g_N}\approx 8Nw^4L^2.
\eeq

The asymptotic behavior of the conductance in the opposite insulating regime ($w^2L\gg1$)
can be investigated as follows.
Typically, i.e., for most realizations of disorder,
the result~(\ref{gsum}) is dominated by the best conducting chain.
We thus have
\beq
g_N\approx 4\,\e^{-2\abs{X_\min(\tau)}},
\eeq
where $\abs{X_\min(\tau)}$ is the smallest absolute value of the $N$ Brownian motions.
The latter quantity reads
\beq
\abs{X_\min(\tau)}=2w\sqrt{L}\,z_N,
\eeq
and so
\beq
\ln g_N\approx\ln4-4w\sqrt{L}\,z_N,
\label{lngsca}
\eeq
where $z_N$ is the smallest of $N$ independent random numbers $x_n$,
drawn from the normalized half-Gaussian law
\beq
f(x)=\frac{2}{\sqrt\pi}\,\e^{-x^2}\quad(x>0).
\eeq
The distribution of $z_N$ can be obtained by means of a standard argument
from extreme-value statistics~\cite{feller}.
We have $\prob{x_n>x}=\erfc x$, where
\beq
\erfc x=\frac{2}{\sqrt\pi}\int_x^\infty\e^{-x'^2}\,\d x'
\eeq
is the complementary error function,
and so $\prob{z_N>z}=(\erfc z)^N$, and
\beq
f(z_N)=\frac{\d}{\d z_N}(\erfc z_N)^N
=\frac{2N}{\sqrt\pi}\;\e^{-z_N^2}(\erfc z_N)^{N-1}.
\label{fzed}
\eeq

The scaling behavior of the distribution of $\ln g_N$ for typical long samples
is readily obtained by inserting the distribution~(\ref{fzed})
into the formula~(\ref{lngsca}):
\beqa
f(\ln g_N)&\approx&\frac{N}{\sqrt{4\pi w^2L}}
\,\exp\!\left(-\frac{(\ln4-\ln g_N)^2}{16w^2L}\right)
\nonumber\\
&\times&\left(\erfc\frac{\ln4-\ln g_N}{4w\sqrt{L}}\right)^{N-1}.
\label{fres}
\eeqa
We have in particular
\beq
\dis{\ln g_N}\approx\ln4-C_N w\sqrt{L},\quad
\var{\ln g_N}\approx D_N w^2L,
\label{gmom}
\eeq
with
\beq
C_N=4\,\dis{z_N},\quad D_N=16\left(\dis{z_N^2}-\dis{z_N}^2\right)
\label{cdres}
\eeq
and
\beq
\dis{z_N}=\int_0^\infty(\erfc z)^N\,\d z,\quad
\dis{z_N^2}=\int_0^\infty2z(\erfc z)^N\,\d z.
\label{zmom}
\eeq
The latter expressions have been derived from~(\ref{fzed})
by means of integrations by parts.

The formulas~(\ref{fres}) and~(\ref{gmom})
constitute the main results of this section.
It is clear from their derivation
that they hold for all values of the disorder strength $w$.
We thus predict that the typical zero-energy conductance $g_N$ of a disordered sample
decays as $\exp(-C_N w\sqrt{L})$ (i.e., subexponentially) with its length~$L$.
Moreover, the conductance distribution is far from the conventional log-normal form
suggested by the one-parameter scaling theory of Anderson localization.
Indeed the relative variance of $\ln g_N$ approaches a non-trivial constant:
\beq
\frac{\var{\ln g_N}}{\left({\dis{\ln g_N}}\right)^2}
\to K_N=\frac{D_N}{C_N^{\,2}}.
\label{kres}
\eeq

For the first values of $N$, the moments involved in~(\ref{cdres})
can be worked out explicitly.
We thus get
\beq
\matrix{
C_1=\frad{4}{\sqrt\pi},\quad
D_1=8-\frad{16}{\pi},\quad
K_1=\frad{\pi}{2}-1
}
\eeq
and
\beq
\matrix{
C_2=\frad{8-4\sqrt{2}}{\sqrt\pi},\quad
D_2=8-\frad{16}{\pi}(7-4\sqrt{2}),\hfill\cr\ms
K_2=\frad{\pi}{4}(3+2\sqrt{2})-\frad{5}{2}-\sqrt{2}.\hfill\cr\ms
}
\eeq
For wide ribbons ($N\gg1$),
the distribution~(\ref{fzed}) of $z_N$ becomes a narrow exponential:
\beq
f(z_N)\approx\frac{2N}{\sqrt\pi}\,\exp\!\left(-\frac{2Nz_N}{\sqrt\pi}\right).
\eeq
We thus obtain the following limits as $N\to\infty$:
\beq
NC_N\to\sqrt{4\pi},\quad N^2D_N\to4\pi,\quad K_N\to1.
\label{cdklim}
\eeq

\begin{figure}[!ht]
\begin{center}
\includegraphics[angle=-90,width=.45\linewidth]{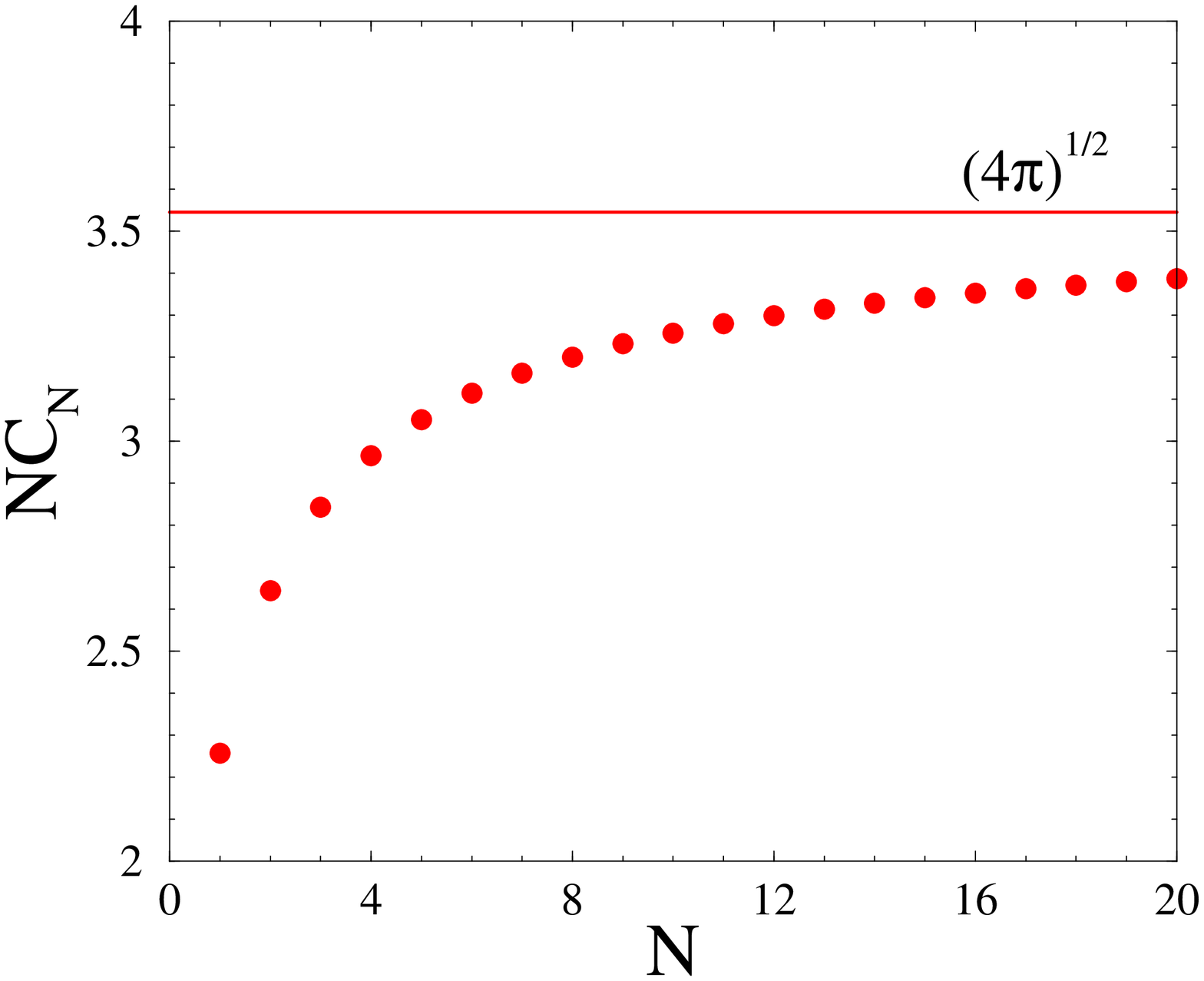}
\includegraphics[angle=-90,width=.45\linewidth]{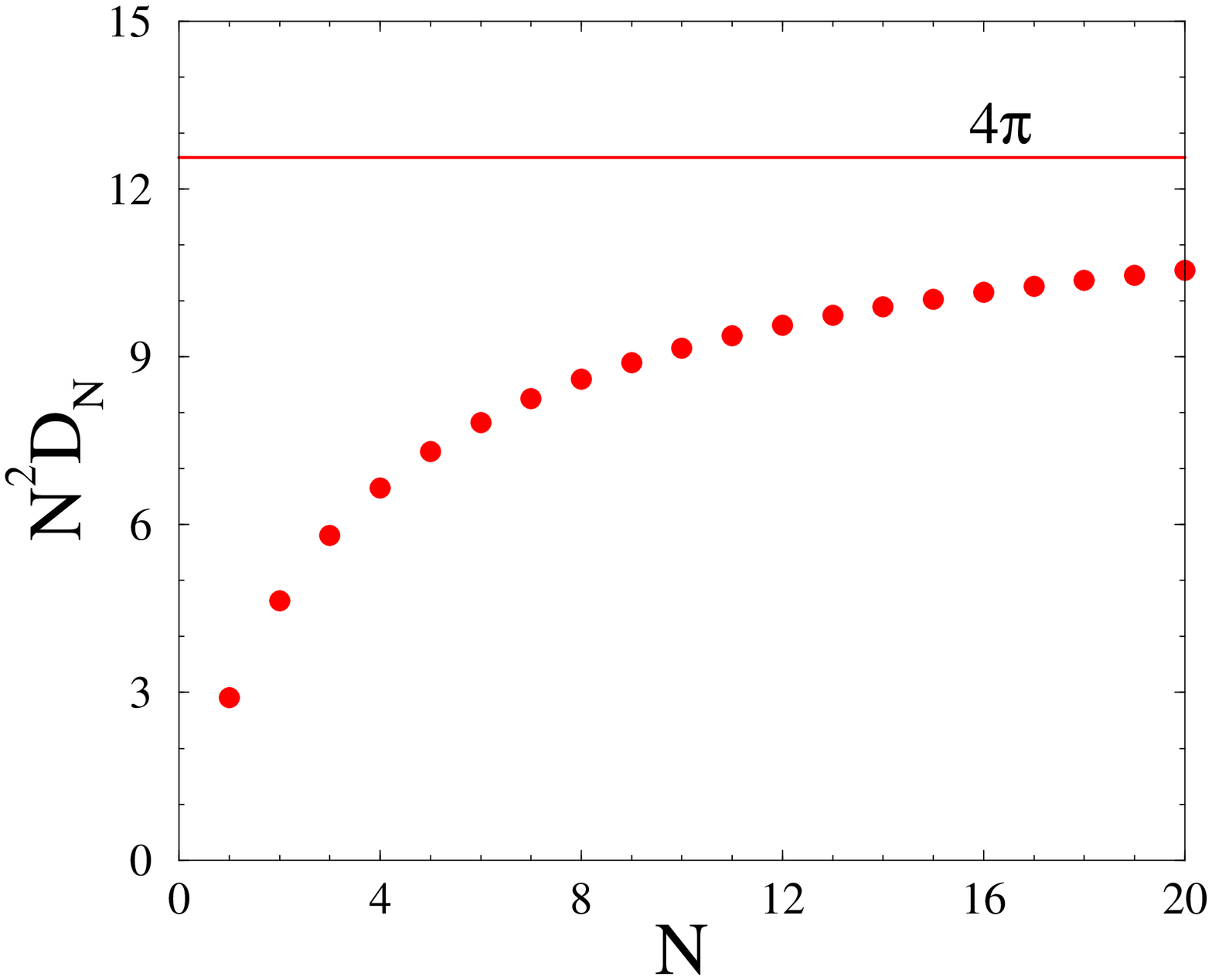}
{\vskip -6pt}
\includegraphics[angle=-90,width=.45\linewidth]{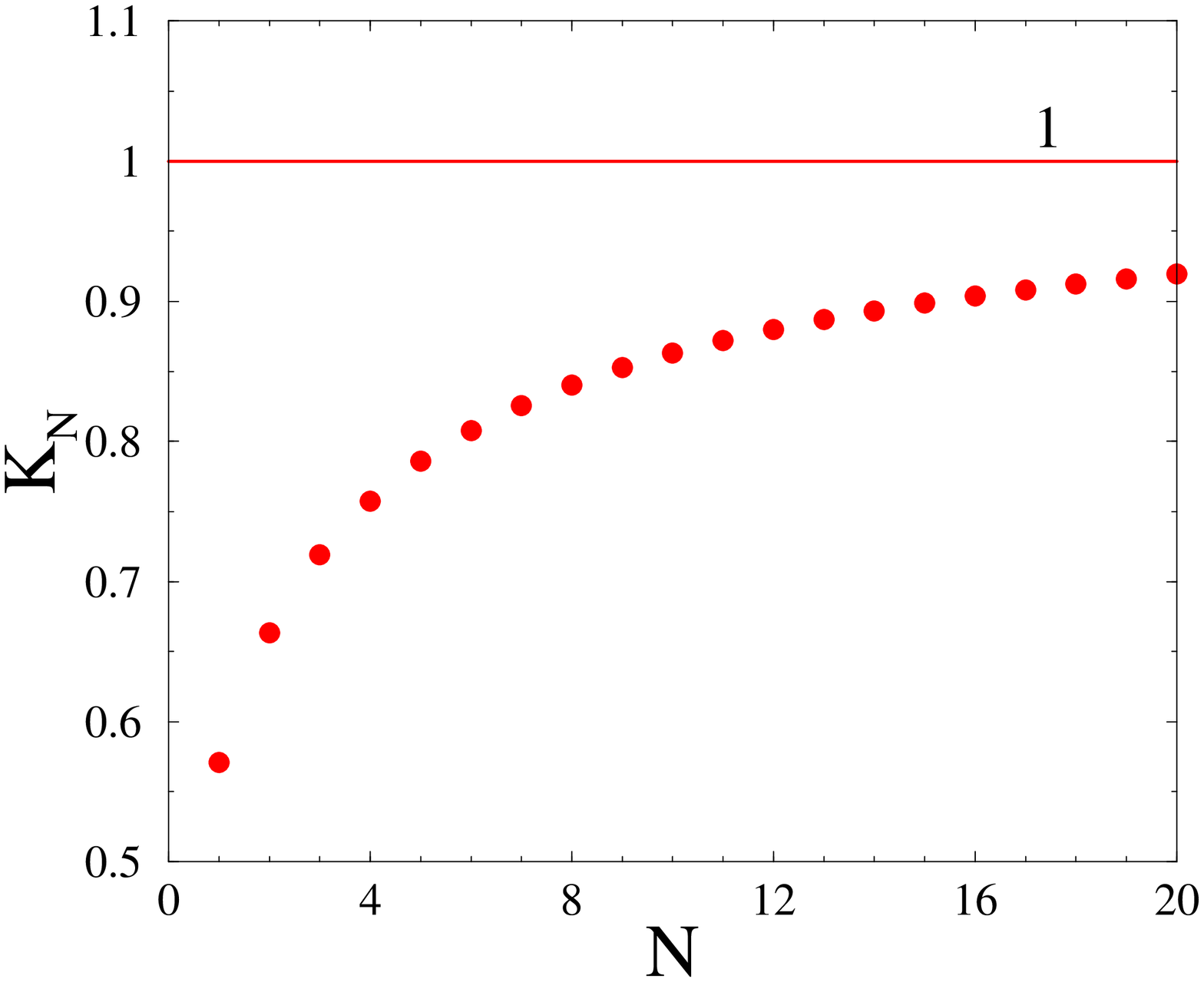}
\caption{\small
Plots of the constants $NC_N$, $N^2D_N$ and $K_N$ against ribbon width~$N$.
Horizontal lines: asymptotic limits~(\ref{cdklim}).}
\label{cdk}
\end{center}
\end{figure}

\begin{figure}[!ht]
\begin{center}
\includegraphics[angle=-90,width=.45\linewidth]{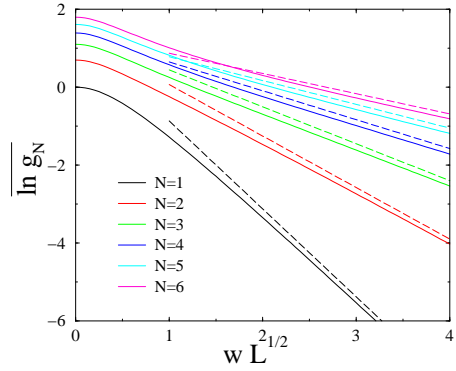}
\caption{\small
Plot of $\dis{\ln g_N}$ against $w\sqrt{L}$,
for ribbon widths $N=1$ to 6 (bottom to top).
Straight dashed lines: exact asymptotic behavior~(\ref{gmom}).}
\label{cl}
\end{center}
\end{figure}

Figure~\ref{cdk} shows plots of $NC_N$, $N^2D_N$ and $K_N$
against the ribbon width $N$.
Horizontal lines show the limits~(\ref{cdklim}).
Figure~\ref{cl} shows a plot of $\dis{\ln g_N}$ against $w\sqrt{L}$,
for ribbon widths up to $N=6$.
Each data point has been obtained by averaging the conductance $g_N$,
as given by the universal expression~(\ref{gsum}), over $10^7$ realizations of disorder.
Straight dashed lines show the asymptotic result~(\ref{gmom}).
All the curves for $N\ge3$ present an inflection point.

\begin{figure}[!ht]
\begin{center}
\includegraphics[angle=-90,width=.45\linewidth]{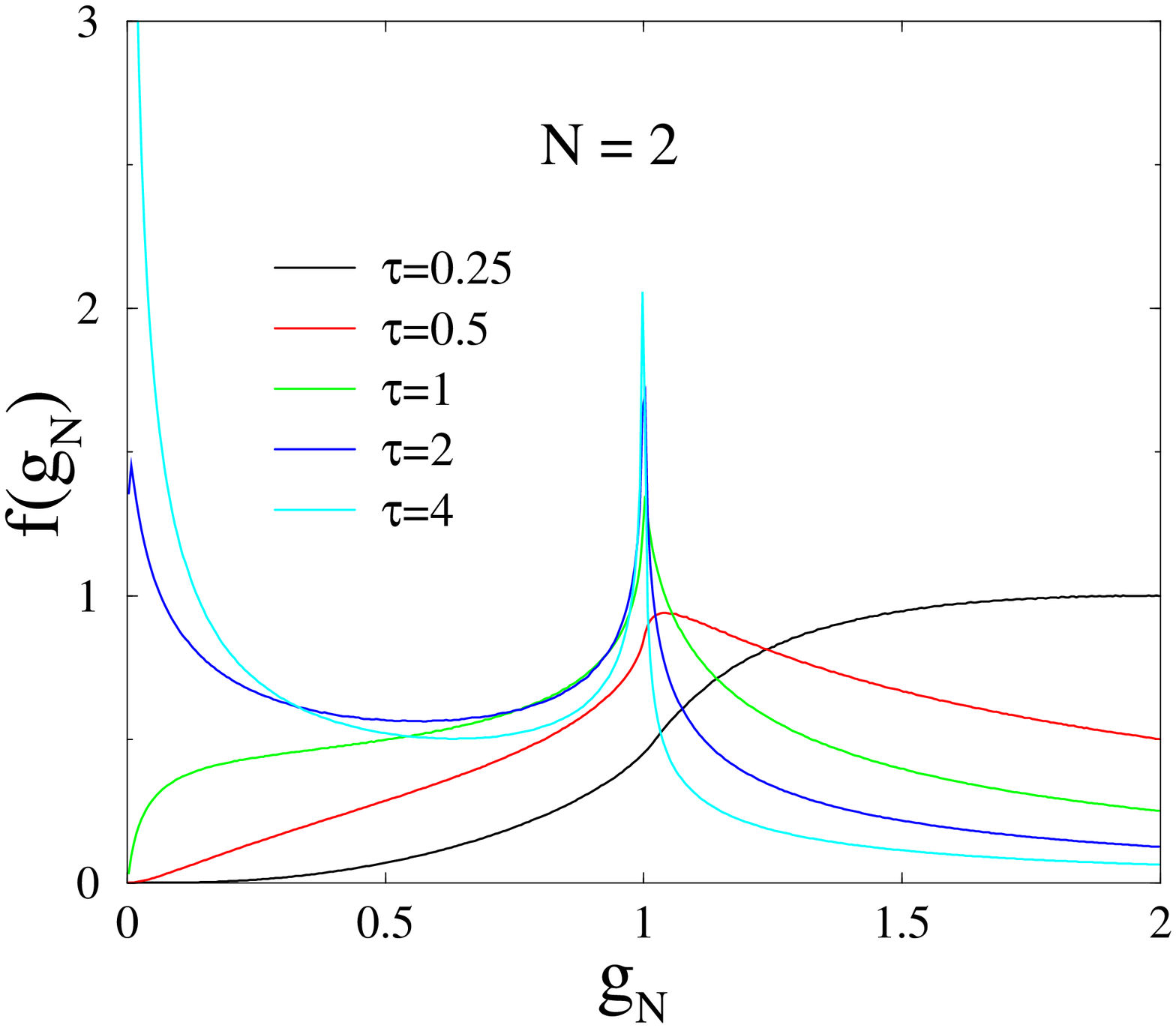}
\includegraphics[angle=-90,width=.45\linewidth]{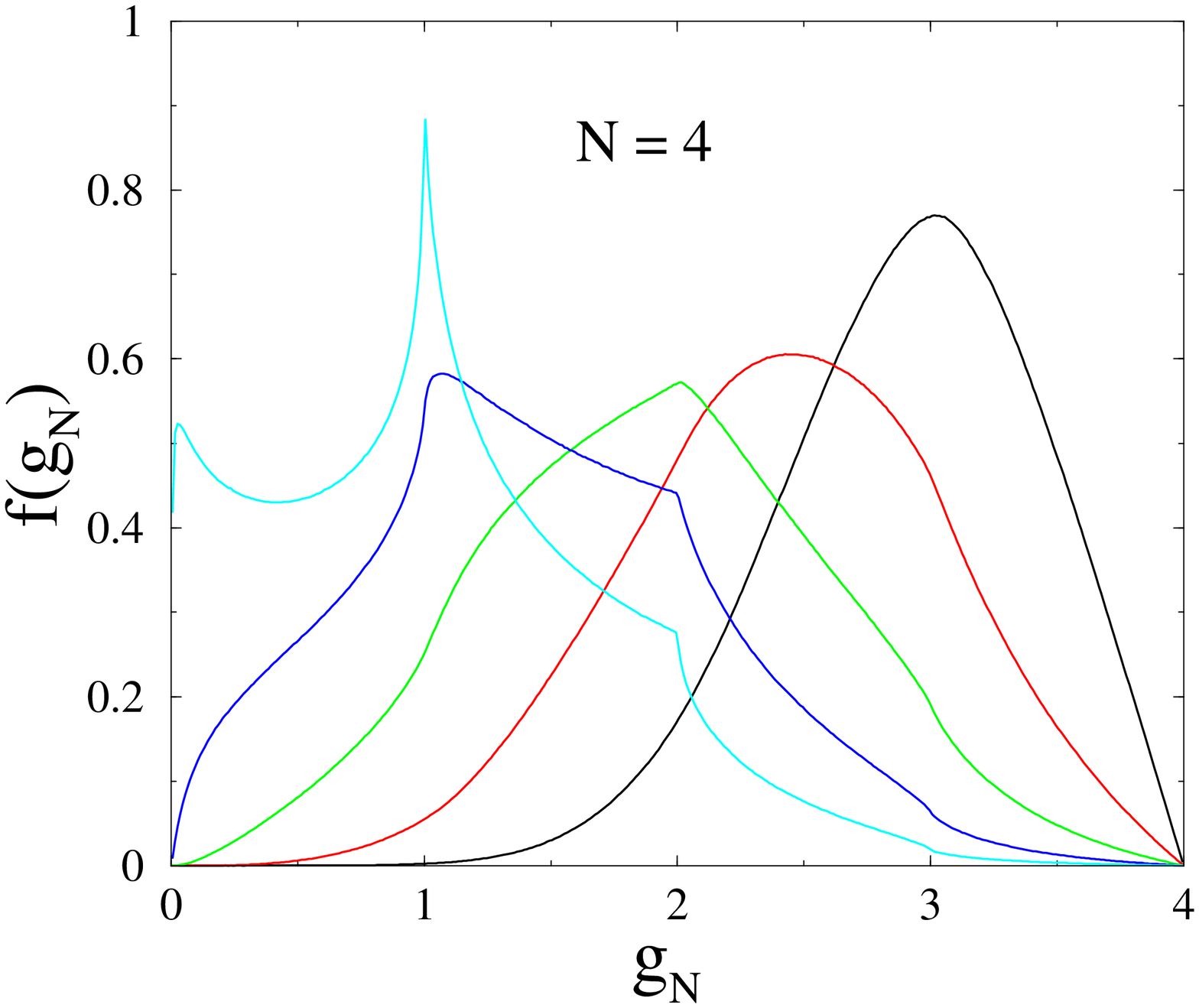}
\caption{\small
Distribution $f(g_N)$ of the zero-energy conductance
for widths $N=2$ and $N=4$ and several values of $\tau=w^2L$
(the color code is common to both panels).}
\label{clin}
\end{center}
\end{figure}

\begin{figure}[!ht]
\begin{center}
\includegraphics[angle=-90,width=.45\linewidth]{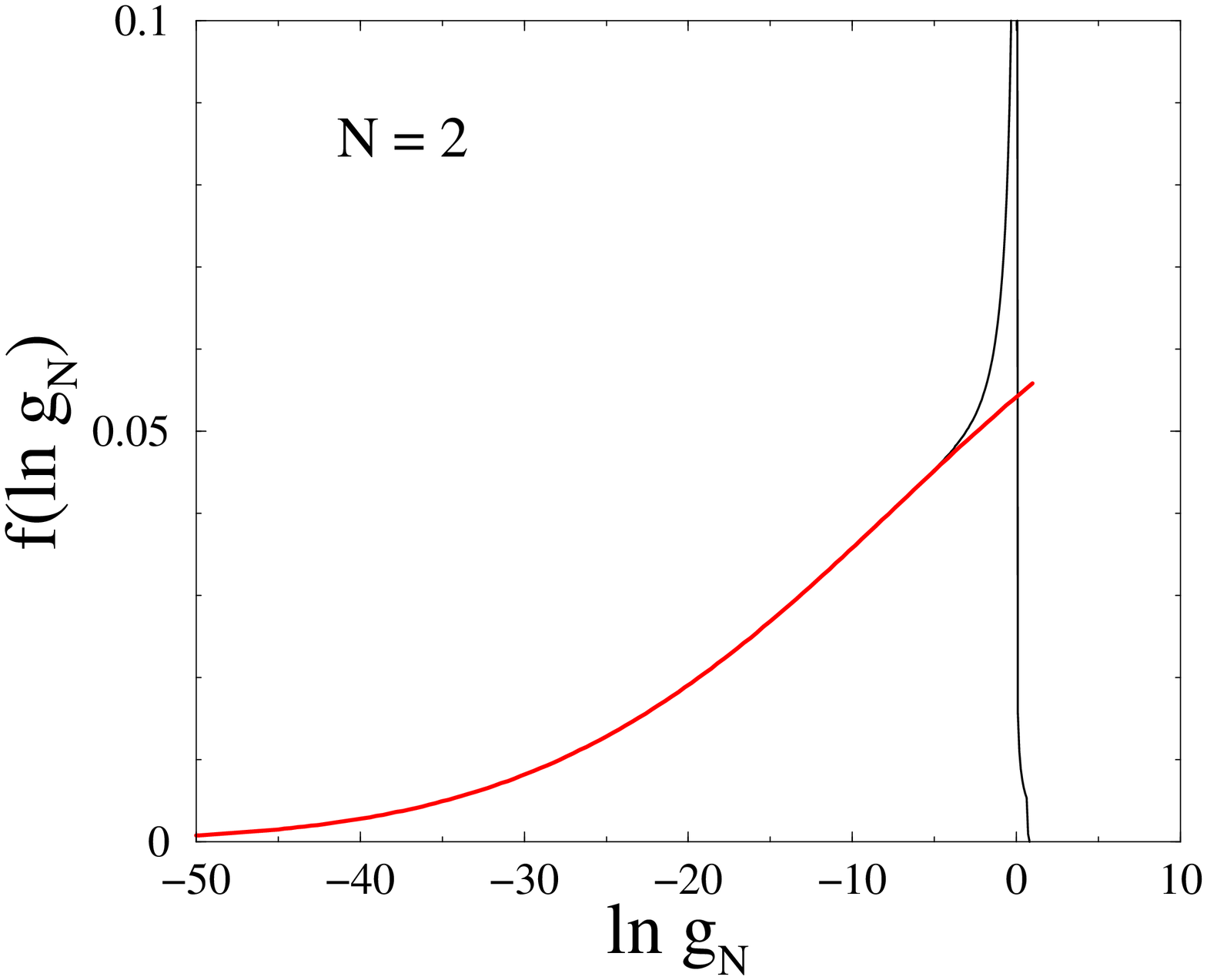}
\includegraphics[angle=-90,width=.45\linewidth]{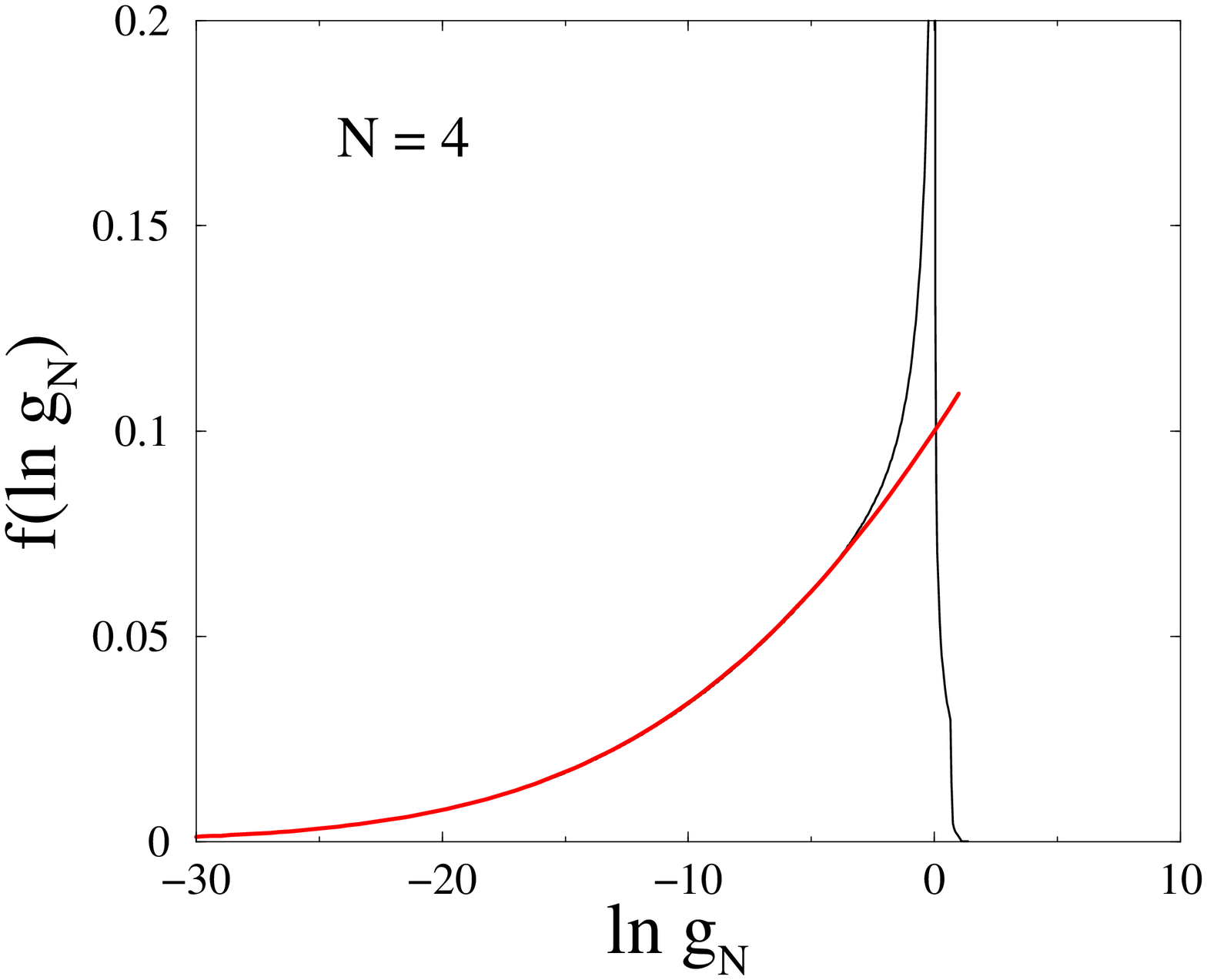}
\caption{\small
Distribution of $\ln g_N$ for $N=2$ and $N=4$ and $w^2L=100$.
Black curves: histograms.
Red curves: analytical prediction~(\ref{fres}).}
\label{csca}
\end{center}
\end{figure}

We close with a few illustrations
of the full distribution of the zero-energy conductance $g_N$,
as given by the universal expression~(\ref{gsum}).
The forthcoming histograms have been obtained
from $10^7$ in\-de\-pen\-dent realizations of disorder.
Fi\-gure~\ref{clin} shows the distribution of $g_N$
for $N=2$ and $N=4$ and several values of $\tau=w^2L$.
Near the maximal ballistic conductance $(g_N\to N)$,
the quadratic approximation~(\ref{quadra}) holds,
and so the distribution has the power-law behavior of~(\ref{gbal}), i.e.,
\beq
f(g_N)\approx\frac{(N-g_N)^{(N-2)/2}}{\Gamma(N/2)(4w^2L)^{N/2}}.
\eeq
This explains why the plotted distributions have finite limiting values for $N=2$,
while they vanish linearly for $N=4$.
The conductance distributions also exhibit internal singularities
at integer values of the conductance ($g_N=1,2,\dots$),
in correspondence with channel thresholds.
Figure~\ref{csca} shows the distribution of $\ln g_N$
for $N=2$ and $N=4$ and $w^2L=100$.
The histograms (black curves) consist of a smooth background,
which is reproduced to a very high accuracy
by our analytical prediction~(\ref{fres}) (red curves),
and of a narrow peak corresponding to $g_N=1$.
The height of this peak exceeds the vertical scale of the plot,
while the peaks at higher integer values of $g_N$ are not visible.

\section{Discussion}
\label{discussion}

The present paper comes after many earlier works devoted to the electronic properties
of zigzag nanoribbons~\cite{zz,fwn,nf,bf,me,pg,kh,ab,eskh,wty,wsn,dum,GM,bdj,dbj,kag}.
It is therefore useful to sum up our main findings.

As far as clean ribbons are concerned (Section~\ref{clean}),
our main point has been to emphasize the role
of the unusual power-law dispersion of the central bands.
The wings of these bands have been known for long
to consist of localized edge states.
The associated power-law dispersion~(\ref{epower}),
with its exponent equal to the width $N$ of the ribbon,
seems however to have been noticed so far only in~\cite{GM}.
This scaling result has several far reaching consequences.
It is in particular responsible for the strong power-law divergence~(\ref{rhopower})
of the density of states near zero energy,
and for the vanishing of the localization length of the edge states~(\ref{xipower}).
Another characteristic property of clean ribbons is the existence
of only two zero-energy Bloch states~(\ref{bloch}), irrespective of the ribbon width.
A related matter concerns the structure
of generic zero-energy eigenstates in a semi-infinite geometry.
These eigenstates grow as various powers of the distance
to the end of the ribbon (see~(\ref{power})).

For a finite ribbon of length $L$,
these zero-energy features are only valid in a tiny neighborhood of zero energy,
defined by the condition $Q_0L\ll1$,
where the momentum scale reads $Q_0=\abs{E}^{1/N}$ (see~(\ref{q0})).
In the situation where the numerical calculations reported in~\cite{kag}
have been performed, i.e., $N=8$ and $E=10^{-6}$,
we predict a surprisingly sizeable momentum $Q_0=0.18$,
in spite of the very small nominal value of energy.
The system is therefore effectively far from zero energy
as soon as its length~$L$ exceeds the very modest length of $1/Q_0\sim 6$ units,
shorter than its width.
This may explain why the beauties described in the present work
were not unveiled by the numerical approach used in~\cite{kag}.

In the presence of hopping (off-diagonal) disorder,
which respects the lattice chiral symmetry (Section~\ref{disorder}),
our leitmotiv is that all zero-energy localization properties are anomalous,
for arbitrary values of the disorder strength $w$.
As a matter of fact,
all observables which would either grow or decay exponentially
in the case of conventional Anderson localization
exhibit an anomalous subexponential behavior on disordered zigzag ribbons
at zero energy.
This holds especially for typical wavefunctions
of generic eigenstates in a semi-infinite geometry,
which grow exponentially as a function of $w\sqrt{l}$ (see~(\ref{abres})),
and for the typical conductance of finite disordered samples of length $L$,
which falls off exponentially as a function of $w\sqrt{L}$ (see~(\ref{gmom})).
These exact scaling results involve constants $A_n$, $B_n$, $C_N$ and $D_N$
which depend only on the channel number $n$ or on the ribbon width~$N$.
We have derived the exact constants $C_N$ and~$D_N$
by means of extreme-value statistics.
It would be desirable to find out an analytical way of deriving
the asymptotic behavior~(\ref{absca}) of the constants $A_n$ and $B_n$.

These findings can be put in perspective with zero-energy properties
of disordered strips made of $N$ coupled channels~\cite{strip1,strip2,strip3,strip4}.
In the latter case, with off-diagonal disorder,
localization properties have been shown to be conventional whenever $N$ is even,
and unconventional whenever $N$ is odd, with subexponential scaling properties
germane to those we have found on zigzag ribbons.
Ribbons and strips could be expected to have some common features.
Indeed the situation of a single chain, investigated by elementary means in~\ref{appa},
can be recovered as the particular case $N=1$ of both ribbons and strips.
The plenty of analytical results derived in this paper however testifies
that zigzag ribbons lend themselves to many exact calculations,
which would just not be possible in other geometries, including strips or armchair ribbons.

Finally, from the viewpoint of the transfer-matrix formalism,
disordered zigzag ribbons provide an interesting case study with many unusual features.
The transfer matrix $T_0$,
describing electron propagation through one unit cell of a clean ribbon at zero energy,
is not diagonalizable.
This property originates in the existence of only two zero-energy Bloch states,
irrespective of the ribbon width.
These features have a remarkable consequence, namely the power-law growth
of generic zero-energy eigenstates in a semi-infinite geometry.
In the presence of hopping (off-diagonal) disorder,
the elements of the matrix products $\T$,
describing propagation through whole disordered samples,
exhibit an anomalous subexponential growth,
related to the similar growth of generic eigenstates
and to the subexponential falloff of the conductance.
It is worth recalling that, in a situation exhibiting conventional Anderson localization,
the matrix product $\T$ would grow as $\exp(\gamma L)$,
with $\gamma>0$ being the largest Lyapunov exponent of the matrix product.
In the present case of zigzag ribbons with off-diagonal disorder at zero energy,
the subexponential growth of $\T$ implies that $\gamma$ vanishes.
To sum up, we have thus explicitly constructed non-trivial random matrix products
such that all Lyapunov exponents vanish identically.

\ack

It is a pleasure to thank Cristina Bena,
Jean-No\"el Fuchs and Gilles Montambaux for fruitful discussions.
The research of YA is partially supported by grant 400/2012
of the Israeli Science Foundation.

\appendix

\section{Zero-energy properties of a disordered chain}
\label{appa}

\begin{figure}[!ht]
\begin{center}
\includegraphics[angle=-90,width=.45\linewidth]{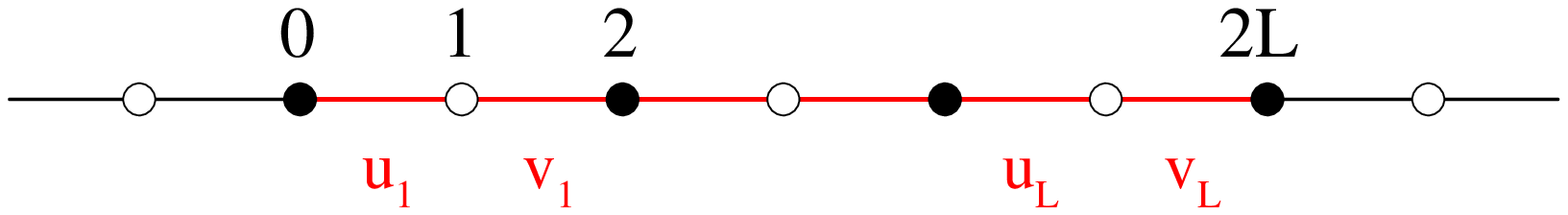}
\caption{\small
A disordered segment embedded in a clean chain.
Black lines: clean bonds with unit hopping amplitudes ($t_n=1$).
Red lines: disordered bonds with random hopping amplitudes ($t_n=u_l$ or $v_l$).}
\label{chain}
\end{center}
\end{figure}

In this Appendix we study by elementary means
various zero-energy properties of a tight-binding chain
with off-diagonal disorder~\cite{TC,FL,SJ,JV}.
We consider the Hamiltonian
\beq
\H=\sum_n t_n(a^\dag_{n-1}a_n+a^\dag_na_{n-1})
\eeq
on a disordered segment of length $2L$
embedded in a clean chain, as shown in Figure~\ref{chain}.

We parametrize the hopping amplitudes of the disordered segment as
\beq
t_{2l-1}=u_l=\exp\bigl(\eps_l^{(u)}\bigr),\quad
t_{2l}=v_l=\exp\bigl(\eps_l^{(v)}\bigr),
\eeq
where the $\eps_l^{(u,v)}$ are again independent random variables,
drawn from an unspecified symmetric probability distribution
such that $\dis{\eps_l}=0$ and $\var{\eps_l}=w^2$.
The disordered sample is connected to two clean semi-infinite leads ($t_n=1$).

\subsection*{Wavefunction}

At zero energy, the amplitudes of the wavefunction at even sites (full symbols)
and odd sites (empty symbols) of the disordered sample decouple and obey the recursions
\beq
\matrix{
u_l\p_{2l-2}+v_l\p_{2l}=0,\hfill\cr\ms
v_l\p_{2l-1}+u_{l+1}\p_{2l+1}=0.\hfill
}
\label{tc}
\eeq
These amplitudes therefore read~\cite{TC,FL,SJ,JV}
\beq
\psi_{2l}=(-1)^l\frac{U_l}{V_l}\p_0,\quad
\psi_{2l+1}=(-1)^l\frac{u_1}{u_{l+1}}\frac{V_l}{U_l}\p_1
\label{puv}
\eeq
in terms of the initial values $\p_0$ and $\p_1$,
where we have introduced the products
\beq
U_l=\prod_{k=1}^l u_k,\quad
V_l=\prod_{k=1}^l v_k.
\label{uvdef}
\eeq

\subsection*{Reflection, transmission, and conductance}

In order to calculate the reflection and transmission amplitudes $r$ and $t$
of the disordered sample at zero energy,
we impose the following wavefunction in the leads
\beq
\p_n=\left\{\matrix{
\e^{\ii qn}+r\,\e^{-\ii qn} & (n\le0),\hfill\cr\ms
t\,\e^{\ii q(n-2L)}\hfill & (n\ge2L),\hfill
}\right.
\eeq
describing the scattering of a particle incoming from the left with momentum $q$
(at zero energy, we have $q=\pi/2$).

First, using the second line of~(\ref{tc}) for $l=0$ and $l=L$,
we obtain the four boundary amplitudes:
\beqa
&&\p_0=1+r,\quad\p_1=\frac{\ii(1-r)}{u_1},\nonumber\\
&&\p_{2L-1}=-\frac{\ii t}{v_L},\quad\p_{2L}=t.
\eeqa

Then, using~(\ref{puv}), we can express the reflection and transmission amplitudes
in terms of the products $V_L$ and $U_L$:
\beq
r=\frac{V_L^2-U_L^2}{V_L^2+U_L^2},\quad
t=(-1)^L\frac{2V_LU_L}{V_L^2+U_L^2}.
\label{rtuv}
\eeq
These reflection and transmission amplitudes are real.
They obey $r^2+t^2=1$, as should be.
The reflection amplitude $r$ vanishes if the equality
\beq
U_L=V_L
\eeq
is fulfilled.
This is the condition on the disorder realization
for the disordered segment of length $2L$ to admit an exact zero-energy eigenstate
with either periodic (if $L$ is even) or anti-periodic (if $L$ is odd) boundary conditions.
We have then $t=(-1)^L$.

Introducing the quantity
\beq
X_L=\ln\frac{V_L}{U_L}=\sum_{l=1}^L\ln\frac{v_l}{u_l}
=\sum_{l=1}^L\left(\eps_l^{(v)}-\eps_l^{(u)}\right),
\label{xuv}
\eeq
the expressions~(\ref{rtuv}) can be recast as
\beq
r=\tanh X_L,\quad t=\frac{(-1)^L}{\cosh X_L}.
\eeq
The dimensionless conductance
of the disordered segment at zero energy is therefore exactly given by
\beq
g=t^2=\frac{1}{\cosh^2 X_L}.
\label{gchain}
\eeq

In the regime of long samples ($L\gg1$),
the rightmost side in~(\ref{xuv}) is the sum of~$2L$ independent random variables
with zero mean and variance $w^2$.
Hence $X_L$ behaves asymptotically
as a Gaussian variable with zero mean and variance $2\tau$,
where we have introduced the shorthand
\beq
\tau=w^2L.
\eeq
The distribution of $X_L$ therefore reads
\beq
f(X_L)=\frac{\e^{-X_L^2/(4\tau)}}{\sqrt{4\pi\tau}}.
\label{xdis}
\eeq
The distribution of the conductance $g$
of a long disordered segment can be obtained by changing variables
from $X_L$ to $g$ in~(\ref{xdis}).
We thus obtain the following universal asymptotic result:
\beq
f(g)=\frac{1}{4g\sqrt{\pi\tau(1-g)}}\,
\exp\!\left(\!-\frac{1}{4\tau}\left(\ln\frac{1+\sqrt{1-g}}{\sqrt{g}}\right)^{\!2}\right),
\label{fgchain}
\eeq
where the disorder strength only enters through $\tau$.

\subsection*{Transfer-matrix formalism}

Our last goal is to check that the result~(\ref{gchain})
is correctly predicted by the transfer-matrix formalism
exposed in Section~\ref{disorder:conductance}.
In the present case, it is advantageous to define the transfer matrix $T_n$
associated with bond number $n$ at zero energy as follows:
\beq
\pmatrix{t_{n+1}\p_{n+1}\cr\p_n}=T_n\pmatrix{t_n\p_n\cr\p_{n-1}}.
\eeq
We have explicitly
\beq
T_n=\pmatrix{0&-t_n\cr1/t_n&0}.
\eeq
The transfer matrices thus defined have unit determinant,
and they are statistically independent from each other.
The transfer matrix $\T$ describing the propagation across
the whole disordered sample reads
\beq
\T=T_{2L}\dots T_1=(-1)^L\pmatrix{\e^{X_L}&0\cr 0&\e^{-X_L}},
\eeq
where $X_L$ has been introduced in~(\ref{xuv}).

The transfer matrix $T_0$ associated with one bond of the leads
and a matrix $\R$ of right eigenvectors of $T_0$ read
\beq
T_0=\pmatrix{0&-1\cr 1&0},\quad
\R=\pmatrix{\ii&-\ii\cr 1&1}.
\eeq
We have therefore
\beq
\M=\R^{-1}\T\R=(-1)^L\pmatrix{\cosh X_L&-\sinh X_L\cr-\sinh X_L&\cosh X_L}.
\eeq
The matrix $\M$ thus obtained is real symmetric.
The eigenvalues of $\M^\dag\M$ are therefore the squares of those of $\M$,
i.e., $\e^{\pm2X_L}$.
Finally,~(\ref{gdef}) and~(\ref{xdef}) yield the expected result~(\ref{gchain}).

\section{Statistics of zero-energy eigenstates at chain number $n=2$}
\label{appb}

In this Appendix we calculate the amplitudes $A_2$ and $B_2$ (see~(\ref{ab2}))
which govern the statistics of zero-energy eigenstates
on a disordered ribbon for $n=2$.

Let us introduce the following linear combinations of $X_1(\tau)$ and $X_2(\tau)$:
\beq
x(\tau)=X_1(\tau)-X_2(\tau),\quad
y(\tau)=X_1(\tau)+X_2(\tau).
\eeq
The latter processes are independent Brownian motions,
so that $\dis{x^2(\tau)}=\dis{y^2(\tau)}=4\tau$.
Moreover, for $n=2$,~(\ref{pln}) and~(\ref{mrec}) read
\beq
M_2(\tau)=m(\tau)=\comport{\max}{0\le\tau'\le\tau}x(\tau')
\eeq
and
\beq
\ln\abs{\p_{4,l}}\approx m(\tau)+\frac12(y(\tau)-x(\tau)).
\label{pln2}
\eeq

The joint distribution of the final position $x(\tau)$ and of the maximum $m(\tau)$
for a single Brownian motion starting at the origin
can be determined by the method of images~\cite{feller}.
For an unrestricted Brownian motion starting at a generic position $x_0$ at time $t=0$,
the probability density of being at position $x$ at time $t$ is given by the free
Green's function
\beq
G(x,\tau;x_0)=\frad{\e^{-(x-x_0)^2/(8\tau)}}{\sqrt{8\pi\tau}}.
\eeq
If the Brownian motion is constrained to obey $x(\tau')<m$ for all times $0<\tau'<\tau$,
the probability density is given by the Dirichlet Green's function
\beq
G_D(x,\tau;x_0)=G(x,\tau;x_0)-G(x,\tau;2m-x_0),
\eeq
constructed from the free one by subtracting the contribution
of an image source located at $2m-x_0$.
The joint probability density $f(x,m)$ is obtained as
\beq
f(x,m)=\frac{\partial}{\partial m}\,G_D(x,\tau;0)
=\frac{(2m-x)\,\e^{-(2m-x)^2/(8\tau)}}{\sqrt{32\pi\tau^3}}.
\label{fxmres}
\eeq
This expression holds in the domain defined by the inequalities $m>0$ and $x<m$.

By integrating the above result over either variable,
we obtain the marginal probability density of the other one, namely
\beq
f_x(x)=\frad{\e^{-x^2/(8\tau)}}{\sqrt{8\pi\tau}},\quad
f_m(m)=\frad{\e^{-m^2/(8\tau)}}{\sqrt{2\pi\tau}}\quad(m>0).
\eeq
The maximum $m(\tau)$ is therefore distributed according to the same half-Gaussian law
as the absolute value $\abs{x(\tau)}$ of the last position.
The expression~(\ref{fxmres}) also yields the joint moments
\beq
\dis{m(\tau)}=\sqrt{\frac{8\tau}{\pi}},\quad
\dis{m^2(\tau)}=\dis{x^2(\tau)}=4\tau,\quad
\dis{x(\tau)m(\tau)}=2\tau.
\eeq
Inserting the latter expressions into~(\ref{pln2}), we obtain
\beq
\dis{\ln\abs{\p_{4,l}}}\approx\dis{m(\tau)}=\sqrt{\frad{8\tau}{\pi}},
\eeq
\beq
\dis{(\ln\abs{\p_{4,l}})^2}\approx\dis{m^2(\tau)}-\dis{x(\tau)m(\tau)}
+\frac12\,\dis{x^2(\tau)}=4\tau.\hfill
\label{pln2res}
\eeq
These results yield the amplitudes $A_2$ and $B_2$ announced in~(\ref{ab2}).
The last two terms on the right-hand side of~(\ref{pln2res}) cancel each other,
so that we have $\dis{(\ln\abs{\p_{4,l}})^2}\approx\dis{m^2(\tau)}$.
This unexpected identity extends neither to higher moments
nor to higher $n$.

\section*{References}

\bibliography{final.bib}

\end{document}